%% file: main.tex
\documentclass[a4paper,noarxiv,11pt,onecolumn,unpublished]{quantumarticle}

\pdfoutput=1\relax
\usepackage[utf8]{inputenc}
\usepackage[english]{babel}
\usepackage[T1]{fontenc}
\usepackage{microtype}
\usepackage{url}

\usepackage{mathtools}
\usepackage{amssymb}
\usepackage{bm}
\usepackage{physics}
\usepackage{soul}

\usepackage{graphicx}
\usepackage{subcaption}
\usepackage{booktabs}
\usepackage{longtable}
\usepackage[dvipsnames]{xcolor}
\usepackage{tikz}
\usetikzlibrary{calc}
\usetikzlibrary{arrows.meta}
\tikzset{
    classical arrow/.style={
        dashed,
        -{Stealth[length=3mm,width=2.2mm]}
    }
}
\usepackage{quantikz}

\usepackage{algorithm}
\usepackage{algpseudocode}

\usepackage{placeins}

\makeatletter
\providecommand*{\theHALG@line}{}
\renewcommand*{\theHALG@line}{%
  \thealgorithm.\arabic{ALG@line}%
}
\makeatother

\usepackage[numbers,sort&compress]{natbib}

\usepackage{hyperref}
\hypersetup{
  colorlinks=true,
  allcolors=blue!55!black
}
\usepackage[
  capitalise,
  nameinlink,
  noabbrev
]{cleveref}

\crefname{appendix}{Appendix}{Appendices}
\Crefname{appendix}{Appendix}{Appendices}

\definecolor{pawseyGold}{HTML}{D4C89E}
\definecolor{pawseyBlack}{HTML}{30251C}
\definecolor{pawseySilver}{HTML}{CFD2D2}
\definecolor{pawseyPeachy}{HTML}{ED8F7B}
\definecolor{pawseyDustyBlue}{HTML}{7BA6AD}
\definecolor{pawseySunYellow}{HTML}{FDC46F}

\title{Quantum approximate stochastic combinatorial optimisation}

\makeatletter
\let\qasco@displaytitle\@title
\newcommand{\qasco@titlebreak}{\ifinner\space\else\\\fi}
\patchcmd{\qasco@displaytitle}
  {stochastic combinatorial}
  {stochastic\qasco@titlebreak combinatorial}
  {}
  {\PackageError{qasco}{Unable to set the title line break}{}}
\patchcmd{\@titleatfontsize}
  {\@title}
  {\qasco@displaytitle\par}
  {}
  {\PackageError{qasco}{Unable to apply the title layout}{}}
\makeatother

\author{Edric Matwiejew}
\email{edric.matwiejew@pawsey.org.au}
\affiliation{Pawsey Supercomputing Research Centre, Perth, WA 6152, Australia}

\author{Paulo E. Santos}
\affiliation{PrioriAnalytica, Adelaide, SA 5000, Australia}

\author{Jingbo B. Wang}
\affiliation{Centre for Quantum Information, Simulation and Algorithms, The University of Western Australia, Perth, WA 6009, Australia}

\author{Ugo Varetto}
\affiliation{Pawsey Supercomputing Research Centre, Perth, WA 6152, Australia}

\date{October 16, 2026}

\begin{document}

\maketitle

\begin{abstract}
Two-stage stochastic combinatorial optimisation concerns discrete decisions made before all relevant information is known. An initial decision is made; once uncertainty is resolved, it can be adapted to the realised scenario, at a cost. The aim is to find the initial decision that minimises the upfront cost plus the expected cost of optimal adaptation. The first-stage decision problem is often NP-hard, and computing the expected minimum recourse cost can be $\#\mathrm{P}$-hard for a single first-stage candidate. We introduce Quantum Approximate Stochastic Combinatorial Optimisation (QASCO), a variational quantum-walk ansatz that coherently incorporates scenario-dependent recourse into its dynamics to search over first-stage decisions without explicitly evaluating expected minimum recourse costs. We demonstrate QASCO for MaxCut with stochastic edge weights, obtain a circuit with polynomial per-layer gate complexity, and study the ansatz via exact state-vector simulation. Across the instances considered, QASCO concentrates amplitude more effectively on stochastic-optimal first-stage decisions than a previously proposed two-stage QAOA construction. We further show that the ansatz can be calibrated using a depth-independent number of parameters and applied to unseen instances without further circuit-parameter optimisation, including joint decision--scenario--recourse spaces spanning more than $2^{32}$ configurations. Under a query-cost model, comparison with the tested classical heuristics using sample-average approximations of the expected minimum recourse cost shows that QASCO more reliably reaches globally optimal first-stage solutions while requiring fewer cut evaluations. These results show how to incorporate scenario-dependent recourse into the ansatz dynamics, providing a general framework for variational quantum walks in two-stage stochastic combinatorial optimisation.
\end{abstract}

\section{Introduction}

\begingroup
\setlength{\emergencystretch}{3em}

Combinatorial optimisation in domains such as vehicle routing, energy systems, telecommunications network design, and production scheduling often must be carried out before all relevant input data are known. A decision based on historical or modelled information may need to be adapted once the remaining uncertainties are resolved. Evaluating a candidate decision therefore requires accounting not only for its immediate cost but also for the recourse actions it may demand across a range of possible scenarios \cite{Dantzig1955,BirgeLouveaux1997}. Two-stage stochastic optimisation with recourse provides a standard framework for this setting. An initial decision is made before uncertainty is resolved, incurring a deterministic first-stage cost, while a second-stage cost captures the optimal adaptation after an uncertain scenario is revealed~\cite{Dantzig1955,BirgeLouveaux1997,WalkupWets1967}. 

Two-stage stochastic programs are computationally demanding because evaluating the expected minimum recourse cost for a given initial decision is often $\#\mathrm{P}$-hard, even when the stochastic parameters have independent probability distributions~\cite {DyerStougie2006Complexity,hanasusanto2016comment}. This has motivated a long line of classical approximation strategies, ranging from sample average approximation (SAA), which replaces the exact expectation value by an empirical average over sampled scenarios \cite{KleywegtShapiroHomemdeMello2002SAA}, to machine-learning methods that approximate the recourse value function or guide candidate selection within a heuristic search~\cite{DumouchelleEtAl2022Neur2SP,NairEtAl2018LearningOptimizers,BengioEtAl2020LearningPrimal}. This places an emphasis on the development of methods that avoid evaluating the expected minimum recourse cost exactly---either by \emph{estimating} it more cheaply or by \emph{generating} promising candidates without evaluating it at all.

In quantum computing, the Quantum Approximate Optimisation Algorithm (QAOA)~\cite{Farhi2014QAOA,Cerezo2021VQA} and variants such as the Quantum Walk-based Optimisation Algorithm (QWOA)~\cite{MarshWang2019QWOA,MarshWang2020Combinatorial} are prominent approaches to deterministic optimisation. These algorithms use a classically parameterised alternating ansatz comprised of a phase-separation unitary that encodes solution costs and a mixing unitary that drives transitions between candidate solutions. In QWOA, the mixing unitary is formulated as a continuous-time quantum walk (CTQW) on a graph tailored to the combinatorial structure of the search space. By exploiting correlations between the phase-encoded cost landscape and this graph structure~\cite{MatwiejewPyeWang2023StructuredSearch}, the quantum walk can improve convergence and substantially reduce the overhead of optimising the classical variational parameters \cite{BennettNoakesWang2024Analysis}. 

Recent work has begun to address stochastic optimisation with recourse using QAOA-based methods. Rotello et al. \cite{rotello2024} use digitised quantum annealing, implemented with a QAOA circuit, to approximately optimise second-stage decisions across scenarios, with an auxiliary register encoding the scenario distribution and controlling the second-stage variables. Quantum amplitude estimation is then used to estimate the expected recourse cost of the prepared state, providing a quadratic improvement in estimation-query complexity over classical Monte Carlo. Assuming the scenario-distribution state can be prepared efficiently, the runtime is independent of the number of scenarios, while representing the distribution as a quantum state provides an exponential reduction in space. However, the method evaluates the recourse objective for a classically supplied first-stage candidate rather than generating that candidate. More closely related to the present work, Kuroiwa et al. \cite{kuroiwa2025} combine a quantum generative adversarial network (qGAN) with QAOA in a single variational circuit. The qGAN encodes the scenario distribution, while QAOA jointly optimises the first- and second-stage variables by minimising the average measured first-stage and per-scenario recourse cost.

Here we introduce Quantum Approximate Stochastic Combinatorial Optimisation (QASCO), a variational quantum-walk ansatz for two-stage stochastic optimisation with recourse. The ansatz coherently aggregates amplitudes with phase-encoded scenario-dependent recourse information before applying a quantum walk over the aggregated first-stage solution space. Recourse information thereby influences the ansatz dynamics used to generate first-stage candidates without requiring explicit evaluation of the expected minimum recourse cost. The main contributions of this work are:

\begin{enumerate}
\item We introduce QASCO, a quantum-walk-based ansatz for two-stage stochastic combinatorial optimisation with recourse. Given an efficiently preparable scenario distribution, it uses scenario-dependent recourse to guide first-stage candidate generation without explicitly evaluating expected minimum recourse costs within the circuit.

\item We construct a QASCO quantum circuit for MaxCut with mutually independent edge-weight perturbations drawn from known finite distributions. Under the stated circuit assumptions, its per-layer gate complexity is polynomial.

\item We compare QASCO with an adaptation of the two-stage QAOA architecture of~\cite{kuroiwa2025} using exact state-vector simulations. Across the tested problem ensembles, QASCO more consistently amplifies stochastic-optimal solutions with comparable parameter-optimisation effort.

\item We evaluate QASCO as a candidate generator under a query-cost model using the ansatz-depth-independent parameterisation scheme developed for QWOA in~\cite{BennettNoakesWang2024Analysis}. Calibrated parameters remain effective on new instances within each problem setting without per-instance optimisation. QASCO reaches a globally optimal solution in all $2400$ test runs and has the lowest expected query cost in all eight settings against the tested classical uniform enumeration, greedy and tabu search methods using SAA estimates of the expected minimum recourse cost.
\end{enumerate}

The remainder of the manuscript is structured as follows. \Cref{sec:background} reviews the relevant background on stochastic combinatorial optimisation with recourse, introduces the stochastic MaxCut benchmark, and summarises the QWOA framework. We then introduce QASCO in \cref{sec:ansatz}. \Cref{sec:maxcut_implementation} details its circuit implementation for stochastic MaxCut, including recourse oracle construction and gate complexity analysis. \Cref{sec:methods} describes the numerical study of QASCO applied to the stochastic MaxCut problem, including benchmark ensemble generation, state-vector simulation, convergence metrics, parameter calibration, and the candidate-generator benchmark. In \cref{sec:numerical_results}, we present our numerical results and discuss them in \cref{sec:discussion}. We conclude in \cref{sec:conclusion}.

\endgroup

\section{Background}
\label{sec:background}

\subsection{Stochastic combinatorial optimisation with recourse}
\label{sec:stochastic_co}

Combinatorial optimisation seeks solutions $x = (x_0, \ldots, x_{n-1}) \in X$ that minimise a real-valued cost function over a discrete decision space. Stochastic combinatorial optimisation addresses problems with uncertainty in parameters. One such approach models a stochastic problem in two stages. In the \emph{first stage}, a solution $x \in X$ is selected based on known deterministic information. In the \emph{second stage}, a scenario $\omega \in \Omega$ specifies the realised values of the uncertain parameters. A recourse action $y \in Y(x,\omega)$ may then be taken, incurring a cost that depends on $x$, $y$, and~$\omega$. The goal is to identify first-stage solutions that minimise the sum of the first-stage cost and the expected minimum recourse cost.

\begin{figure}[t]
    \centering
    \includegraphics[width=\linewidth]{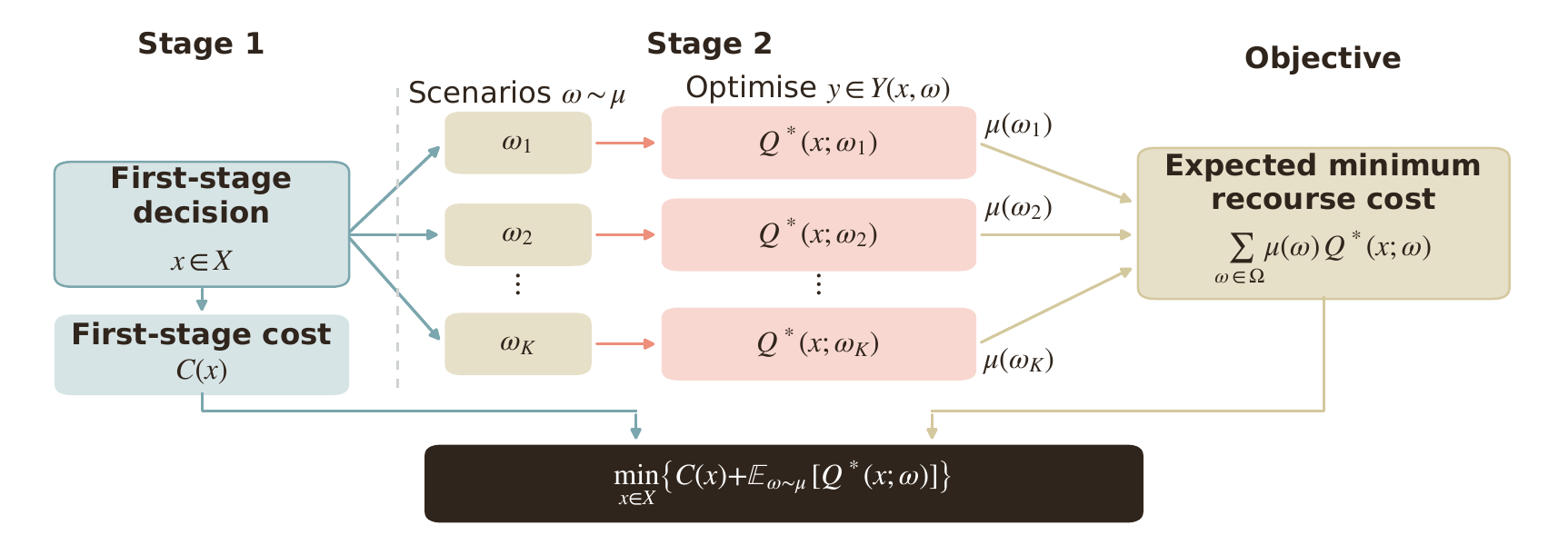}
    \caption{Two-stage stochastic optimisation with recourse. A first-stage decision $x$ is selected before uncertainty is revealed, incurring the deterministic cost $C(x)$. For each scenario $\omega\in\Omega$, the recourse action is chosen from $Y(x,\omega)$ to minimise $Q(x,y;\omega)$, giving $Q^*(x;\omega)$. Averaging this minimum cost under $\mu$ and adding $C(x)$ gives the stochastic objective $J_\mu(x)$.}
    \label{fig:two-stage-recourse}
\end{figure}

More concretely, let $\omega = (\omega_0, \ldots, \omega_{d_\omega-1})$ denote a realisation of $d_\omega$ uncertain parameters, drawn from a finite scenario set $\Omega$ with joint probability mass function $\mu$,
\begin{equation}
    \label{eq:target_scenario_distribution}
    \mu(\omega) \ge 0,
    \qquad
    \sum_{\omega \in \Omega} \mu(\omega) = 1.
\end{equation}
The first-stage cost $C(x)$ is a deterministic function that captures costs that can be evaluated before uncertainty is resolved, independent of $\mu$ \cite{Dantzig1955,BirgeLouveaux1997}. In the second stage, taking recourse action $y \in Y(x,\omega)$ incurs a recourse cost $Q(x,y;\omega)$ that depends on the first-stage decision, the recourse action, and the realised scenario. The minimum recourse cost for a given first-stage decision $x$ and scenario $\omega$ is
\begin{equation}
\label{eq:recourse_function}
    Q^*(x;\omega) = \min_{y \in Y(x,\omega)} Q(x,y;\omega),
\end{equation}
which, together with $C(x)$, defines the two-stage objective function
\begin{equation}
\label{eq:problem_objective}
    J_\mu(x)
    :=
        C(x)
        +
        \mathbb E_{\omega \sim \mu}\big[Q^*(x;\omega)\big],
    \qquad
    \min_{x\in X}J_\mu(x),
\end{equation}
where $\mathbb{E}_{\omega \sim \mu}[Q^*(x;\omega)] = \sum_{\omega \in \Omega}\mu(\omega)\,Q^*(x;\omega)$ is the expectation of the minimum recourse cost over all possible scenarios~\cite{BirgeLouveaux1997,WalkupWets1967}. We illustrate the relationship between $J_\mu(x)$ and the various components of the two-stage stochastic optimisation problem in \cref{fig:two-stage-recourse}.

\subsubsection{Stochastic MaxCut}
\label{sec:maxcut}

\begin{figure}[t]
    \centering
    \includegraphics[width=\linewidth]{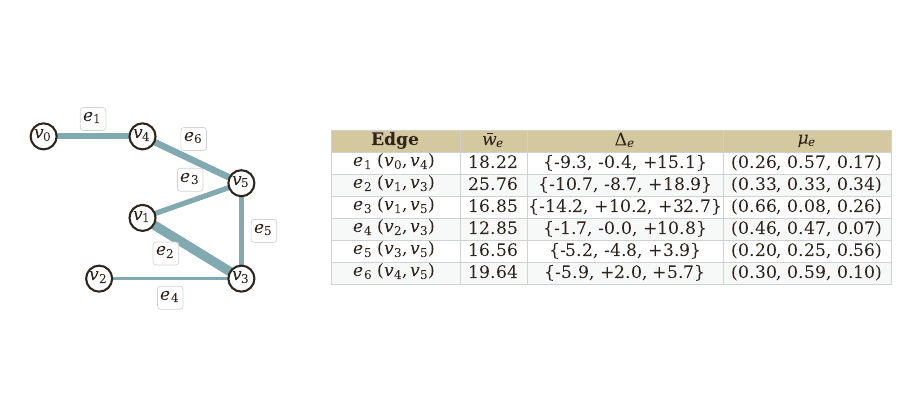}
    \caption{Representative six-vertex stochastic MaxCut instance. \textbf{Left}: graph structure with edge thickness proportional to the nominal weight~$\bar w_e$. \textbf{Right}: nominal weight~$\bar w_e$, perturbation support~$\Delta_e$, and marginal distribution~$\mu_e$ for each edge.}
    \label{fig:random-graph-mean-recourse-instance}
\end{figure}

In this work, we consider the familiar weighted MaxCut problem with stochastic perturbations of the edge weights. The goal is to identify partitions of graph vertices into two sets such that the sum of the nominal edge weights between the sets (the ``cut value'') is large, while the expected minimum recourse cost under stochastic perturbations of the edge weights is small. A cut whose value falls below the target cut value can be repaired, when possible, by moving a minimal number of vertices to the other side of the cut. The resulting repair structure is closely related to recoverable robust optimisation, where an initial solution may be modified after uncertainty is realised \cite{Grune2025HammingRecoverableRobust}, and to two-stage stochastic combinatorial optimisation problems with binary recourse \cite{MacNeilBodur2023BinaryRecourse}. 
Threshold-based models also arise in chance-constrained optimisation and reliable stochastic optimisation \cite{CharnesCooper1959,NikolovaBrandKarger2006,Nikolova2010}, where the resulting non-linearity generally prevents reduction to a deterministic problem instance built on the mean of the uncertain parameters.

Let $G = (V, E)$ be an undirected graph with $|V| = n$ vertices, where each edge $e \in E$ has a known nominal weight $\bar w_e > 0$. A partition $x = (x_0, \ldots, x_{n-1}) \in \{0,1\}^n$ corresponds to spin values $z_i := 1-2x_i \in \{+1,-1\}$, so $z_i \neq z_j$ when vertices $i$ and $j$ lie on opposite sides of the cut. The first-stage cost is the negative nominal cut value,
\begin{equation}
    C(x)
    =
    -\sum_{e = \{u,v\} \in E} \bar w_e\, \frac{1-z_uz_v}{2}.
\end{equation}

Each edge carries a random perturbation $\omega_e$, drawn independently across edges from a zero-mean marginal distribution $\mu_e$ with finite support $\Delta_e$. The scenario space factorises as $\Omega = \prod_{e \in E}\Delta_e$, with one perturbation per edge so that $d_\omega = |E|$. The realised cut value under scenario $\omega$ is
\begin{equation}
    W(x;\omega) = \sum_{e = \{u,v\} \in E} \bigl(\bar w_e + \omega_e\bigr)\,\frac{1-z_uz_v}{2}.
\end{equation}

Under scenario $\omega$, a first-stage cut $x$ requires repair whenever $W(x;\omega) < T$, with a candidate repair $y$ being successful whenever $W(y;\omega) \ge T$. The space of potential recourse actions is $Y(x,\omega) = \{0,1\}^n$. If the original cut falls below the threshold, a successful repair costs the Hamming distance $d_H(x,y)$ (number of bit flips) between the original and repaired partitions, while an unsuccessful repair costs $\Gamma = n + 1$. The recourse cost for recourse action $y$ under scenario $\omega$, given first-stage cut $x$, is
\begin{equation}
\label{eq:Q_piecewise_maxcut}
Q(x,y;\omega)
=
\begin{cases}
0, & W(x;\omega) \ge T, \\[4pt]
d_H(x,y), & W(x;\omega) < T \text{ and } W(y;\omega) \ge T, \\[4pt]
\Gamma, & W(x;\omega) < T \text{ and } W(y;\omega) < T.
\end{cases}
\end{equation}

As a cut and its bitwise complement describe the same partition, we define the canonical domain $X_{\mathrm{can}}:=\{x\in\{0,1\}^n:x_0=0\}$. We denote the sets of canonical cuts minimising the nominal objective $C(x)$ and the stochastic objective $J_\mu(x)$ by
\begin{equation}
    X_C^*:=\operatorname*{arg\,min}_{x\in X_{\mathrm{can}}}C(x),
    \qquad
    X_\mu^*:=\operatorname*{arg\,min}_{x\in X_{\mathrm{can}}}J_\mu(x),
\end{equation}
respectively, and refer to these as the nominal and stochastic optima. We show an illustrative stochastic MaxCut instance in Figure~\ref{fig:random-graph-mean-recourse-instance}.

\subsection{Quantum approximate optimisation} 
\label{sec:qwoa}

The Quantum Approximate Optimisation Algorithm (QAOA) and related ans\"atze provide a standard near-term approach to discrete optimisation on quantum devices \cite{Farhi2014QAOA,Preskill2018NISQ,Cerezo2021VQA}. The QAOA alternates two parameterised unitaries: a \emph{phase-separation} operator $U_C(\gamma) = e^{-i\gamma \hat C}$, where $\hat C$ is the diagonal operator with $\hat C\ket{x} = C(x)\ket{x}$, which imprints the objective function $C$ as relative phases on computational basis states, and a \emph{mixing} operator $U_M(t)$, which redistributes amplitude among candidate solutions. Starting from a uniform superposition, $p$ layers of these operators produce the ansatz state
\begin{equation}
    \ket{\psi(\boldsymbol\gamma,\boldsymbol t)} = U_M(t_p)\,U_C(\gamma_p)\cdots U_M(t_1)\,U_C(\gamma_1)\ket{s},
\end{equation}
where $\boldsymbol\gamma = (\gamma_1,\ldots,\gamma_p)$ and $\boldsymbol t = (t_1,\ldots,t_p)$ are variational parameters optimised classically to minimise the expectation value of the measured solution costs, thereby amplifying low-cost solutions. In the original QAOA formulation, the mixer is the transverse-field unitary $U_M(t) = e^{-it\sum_j \hat X_j}$, where $\hat X_j$ is the Pauli-$X$ operator on qubit $j$.

The Quantum Walk-based Optimisation Algorithm (QWOA) generalises the choice of mixing graph, using a continuous-time quantum walk (CTQW) $U_W(t) = e^{-itA}$ on a problem-specific graph with adjacency matrix $A$~\cite{ManouchehriWang2014QuantumWalks}, whose vertices represent feasible solutions \cite{MarshWang2019QWOA,MarshWang2020Combinatorial}. The QAOA transverse-field mixer is exactly equivalent to a continuous-time quantum walk on the hypercube graph connecting $\{0,1\}^n$. The mixing graph can be chosen to reflect the combinatorial structure and constraints of the feasible solution space, subject to an efficient circuit implementation. Recent constructions define graph edges using the Hamming distance between solutions~\cite{BennettNoakesWang2024NonVariational,BennettNoakesWang2024Analysis}.

Optimising QAOA or QWOA parameters can incur substantial classical overhead, particularly as the circuit depth and number of free parameters increase~\cite{Cerezo2021VQA,zhou_quantum_2020}. This has motivated parameter-initialisation strategies such as INTERP~\cite{zhou_quantum_2020}, which warm-starts depth-$p$ optimisation by linearly interpolating the optimised phase and mixer schedules from depth $p-1$ onto the new layer grid. Recent work on QWOA has introduced a compact linear-ramp parameterisation in which, given an appropriate problem-class-specific mixing graph, the entire depth-$p$ schedule is generated from only three parameters, $\boldsymbol\vartheta:=(\gamma_{\max},t_{\max},\lambda)$~\cite{BennettNoakesWang2024NonVariational,BennettNoakesWang2024Analysis}.  The resulting schedule can achieve convergence to low-cost solutions that is close to that achived by full variational optimisation of the $2p$ ansatsz parameters with fewer evaluations of the ansatz objective function~\cite{MatwiejewEtAl2026VehicleRouting}. Numerical simulations of random weighted-MaxCut instances found that it maintained an approximately constant optimal-solution probability while the required layer count grew quadratically with the number of vertices over the simulated range~\cite{bennett2025benchmarkingquantumheuristicsnonvariational}. Under this parameterisation,
\begin{equation}
\label{eq:linear_ramp}
    \gamma_k = \bigl[\lambda+(1-\lambda)s_k\bigr]\gamma_{\max},
    \qquad
    t_k = \bigl[1-(1-\lambda)s_k\bigr]t_{\max},
\end{equation}
where, for layer $k\in\{1,\ldots,p\}$, $s_k:=(k-1)/(p-1)$ when $p>1$, and $s_1:=0$ when $p=1$.

\section{Variational ansatz for stochastic optimisation}
\label{sec:ansatz}

\begin{figure}[t]
    \centering

    \def\qascooverlay{%
        \coordinate (repeatcentre) at
            ($(\tikzcdmatrixname-2-4.center)!0.5!(\tikzcdmatrixname-2-10.center)$);
        \coordinate (repeatbottom) at
            ($(repeatcentre |- \tikzcdmatrixname-3-4.south)+(0,-0.34cm)$);

        \node[
            draw,
            rounded corners=1pt,
            minimum width=1.35cm,
            minimum height=0.78cm,
            inner sep=2pt,
            font=\ttfamily
        ] (optimizer) at
            ($(repeatcentre |- \tikzcdmatrixname-3-4.south)+(0,-1.55cm)$)
            {>\_};

        \draw
            ($(optimizer.south)+(0,-0.02cm)$)
            -- ++(0,-0.16cm);

        \draw
            ($(optimizer.south)+(-0.22cm,-0.18cm)$)
            --
            ($(optimizer.south)+(0.22cm,-0.18cm)$);

        \coordinate (meastrunk) at
            ($(\tikzcdmatrixname-1-12.east)+(0.85cm,0)$);

        \node[
            inner sep=2pt
        ] (candidate) at
            ($(\tikzcdmatrixname-1-12.east)+(2.4cm,0)$)
            {$\tilde{x}\sim P_X$};

        \node[
            align=center,
            font=\fontsize{11.6}{13}\selectfont
            ] at 
            ($(candidate.north)+(0,0.38cm)$)
            {output\\candidates};

        \draw[classical arrow]
            (\tikzcdmatrixname-1-12.east)
            -- (candidate.west);

        \draw[dashed]
            (\tikzcdmatrixname-2-12.east)
            -- (\tikzcdmatrixname-2-12.east -| meastrunk);

        \draw[dashed]
            (\tikzcdmatrixname-3-12.east)
            -- (\tikzcdmatrixname-3-12.east -| meastrunk);

        \draw[dashed]
            (meastrunk)
            -- (optimizer.east -| meastrunk);

        \draw[classical arrow]
            (optimizer.east -| meastrunk)
            -- (optimizer.east);

        \node[
            align=center,
            font=\fontsize{11.6}{13}\selectfont
        ] at ($
            (optimizer.east -| meastrunk)
            !0.5!
            (optimizer.east)
            +(0,-0.34cm)
        $)
            {estimate $f_{\mathrm{joint}}$};

        \draw[classical arrow]
            (optimizer.north)
            -- (repeatbottom)
            node[midway,right,fill=white,inner sep=3pt,xshift=0.1cm]
            {$\boldsymbol{\theta}' = (\boldsymbol{\gamma}', \boldsymbol{t}') $};
    }

    \resizebox{\linewidth}{!}{%
    \fontsize{13}{15}\selectfont
    \begin{quantikz}[
        row sep={1.30cm,between origins},
        column sep=0.38cm,
        execute at end picture={\qascooverlay}
    ]
        \lstick{$\ket{0}_X$}
            & \gate[style={fill=pawseyDustyBlue!30}]{S_X}
            & \gate[3,style={fill=pawseyDustyBlue!30}]{S_Y}
            & \gate[style={fill=pawseyPeachy!35}]{U_C(\gamma_k)}
              \gategroup[
                  wires=3,
                  steps=7,
                  style={
                      dashed,
                      rounded corners,
                      inner xsep=4pt,
                      inner ysep=7pt
                  },
                  label style={
                      label position=above,
                      yshift=0.18cm
                  }
              ]{repeat $p$ times}
            & \gate[3,style={fill=pawseyPeachy!35}]{U_Q(\gamma_k)}
            & \gate[3,style={fill=pawseyGold!55}]{S_Y^\dagger}
            & \qw
            & \gate[style={fill=pawseySunYellow!50}]{U_W(t_k)}
            & \qw
            & \gate[3,style={fill=pawseyGold!55}]{S_Y}
            & \meter[style={fill=pawseySilver!60}]{x}
            & \qw \\
        \lstick{$\ket{0}_\Omega$}
            & \gate[style={fill=pawseyDustyBlue!30}]{S_\Omega}
            & \qw
            & \qw
            & \qw
            & \qw
            & \gate[
                style={fill=pawseyGold!55},
                wires=1
              ][1.85cm][0.75cm]{S_\Omega^\dagger}
            & \octrl{-1}
            & \gate[
                style={fill=pawseyGold!55},
                wires=1
              ][1.85cm][0.75cm]{S_\Omega}
            & \qw
            & \meter[style={fill=pawseySilver!60}]{\omega}
            & \qw \\
        \lstick{$\ket{0}_{Y}$}
            & \qw
            & \qw
            & \qw
            & \qw
            & \qw
            & \qw
            & \octrl{-2}
            & \qw
            & \qw
            & \meter[style={fill=pawseySilver!60}]{y}
            & \qw
    \end{quantikz}%
    }

    \caption{Schematic circuit for QASCO. $S_X$ prepares the solution register in a uniform state over all first-stage solutions, and $S_\Omega$ prepares a probability-weighted superposition over the space of scenarios according to the target distribution $\mu$. The recourse register is prepared conditionally on the solution and scenario registers by $S_Y$. $U_C$ and $U_Q$ phase-encode the first-stage and recourse costs, respectively. The adjoint preparation unitaries $S_Y^\dagger$ and $S_\Omega^\dagger$ aggregate amplitudes associated with the same first-stage choice in a component flagged by $\ket{0}_\Omega\ket{0}_Y$. The controlled mixer $\Lambda_0(U_W)$ acts on this component before the amplitudes are redistributed back to the full problem configuration space with $S_Y$ and $S_\Omega$. The ansatz objective function is estimated from measurement outcomes, and a classical optimiser updates the variational parameters. Measurements of the solution register produce first-stage candidates, whose stochastic objective values are estimated classically before final selection.}
    \label{fig:stochastic-recourse-ansatz-circuit}
\end{figure}
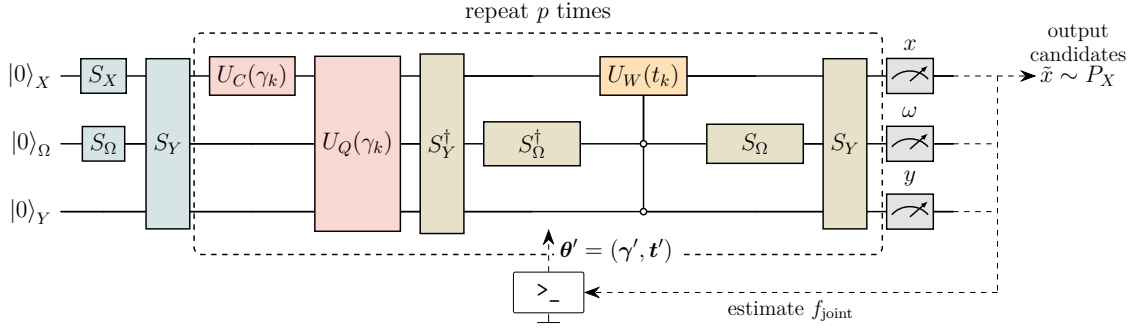

We now introduce QASCO, a variational quantum ansatz for two-stage stochastic optimisation with recourse. The ansatz represents the problem configuration space using registers for the first-stage decision, scenario distribution, and recourse actions, coupling them so that scenario-weighted recourse costs guide amplitude transfer between first-stage candidates. We first define the joint configuration-space encoding and initialisation of the three registers. We then construct a single ansatz layer, specify the variational objective and the role of recourse-state preparation, and finally give a CTQW interpretation of the resulting mixing dynamics. The QASCO quantum circuit is illustrated in \cref{fig:stochastic-recourse-ansatz-circuit}.

\subsection{Configuration-space encoding}

A solution $x \in X$, a scenario $\omega \in \Omega$, and a recourse action $y \in Y$ are respectively encoded in three quantum registers in the computational basis, with Hilbert spaces $\mathcal{H}_X$, $\mathcal{H}_\Omega$, and $\mathcal{H}_Y$. The joint Hilbert space is
\begin{equation*}
\mathcal{H} = \mathcal{H}_X \otimes \mathcal{H}_\Omega \otimes \mathcal{H}_Y,
\end{equation*}
with computational basis states $\ket{x}_X\ket{\omega}_\Omega\ket{y}_Y$.

The scenario Hilbert space decomposes as
\begin{equation*}
    \mathcal{H}_\Omega = \bigotimes_{i=0}^{d_\omega-1} \mathcal{H}_{\Omega_i},
\end{equation*}
where $\mathcal{H}_{\Omega_i}$ is the Hilbert space of the subregister encoding $\Omega_i$. For a scenario $\omega = (\omega_0, \ldots, \omega_{d_\omega-1}) \in \Omega$, the corresponding scenario basis state is
\begin{equation*}
    \ket{\omega}_\Omega = \ket{\omega_0}_{\Omega_0} \cdots \ket{\omega_{d_\omega-1}}_{\Omega_{d_\omega-1}}.
\end{equation*}

\subsection{State initialisation}
\label{sec:state_initialisation}

The ansatz is initialised as a product state of the solution and scenario registers followed by the, potentially controlled, preparation of the recourse register. Let $S_X$ and $S_\Omega$ denote the solution and scenario state-preparation unitaries, satisfying
\begin{align}
    S_X\ket{0}_X &= \ket{s}_X
    :=
    \frac{1}{\sqrt{|X|}}\sum_{x \in X}\ket{x}_X, \\
    S_\Omega\ket{0}_\Omega &= \ket{\mu}_\Omega
    :=
    \sum_{\omega \in \Omega}\sqrt{\mu(\omega)}\,\ket{\omega}_\Omega,
\end{align}
where $\mu$ is the target scenario distribution defined in \cref{eq:target_scenario_distribution}. While the ansatz described in the rest of this section can use any efficiently preparable scenario distribution, including those with correlated uncertain parameters, we focus here on mutually independent parameters, as this allows a concrete analysis of the circuit components for a broad range of practical problems. In this case, we prepare the scenario state using the marginal distributions $\mu_i$:
\begin{equation*}
    S_\Omega
    =
    \bigotimes_{i=0}^{d_\omega-1} S_{\Omega_i},
    S_{\Omega_i}\ket{0}_{\Omega_i}
    = \ket{\mu_i}_{\Omega_i}
    :=
    \sum_{\omega_i \in \Omega_i}\sqrt{\mu_i(\omega_i)}\,\ket{\omega_i}_{\Omega_i},
\end{equation*}
where each unitary $S_{\Omega_i}$ acts on $\lceil\log_2|\Omega_i|\rceil$ qubits (see \cref{sec:appendix_independence_assumption}). Efficient state preparation is available for certain structured distributions~\cite{GroverRudolph2002}, while variational approaches can approximate other target distributions~\cite{Zoufal2019qGAN}. Otherwise, marginal distributions with small discrete support can be prepared by direct unitary synthesis~\cite{Mottonen2004}.

Subsequently, we apply the recourse preparation unitary
\begin{equation}
    S_Y
    =
    \sum_{x\in X}\sum_{\omega\in\Omega}
    \ket{x,\omega}\!\bra{x,\omega}_{X\Omega}
    \otimes S_Y^{(x,\omega)},
\end{equation}
where $S_Y^{(x,\omega)}$ denotes the conditional block acting on $Y$, with $S_Y^{(x,\omega)}\ket{0}_Y=\ket{s_{x,\omega}}_Y$. The normalised state $\ket{s_{x,\omega}}_Y$ defines the reference distribution over recourse actions associated with $(x,\omega)$ and, for recourse spaces with hard constraints, we require $\braket{y}{s_{x,\omega}}=0$ whenever $y\notin Y(x,\omega)$. The challenges associated with recourse preparation and the preparation strategies explored in this work are discussed in \cref{sec:recourse_preparation}.

Altogether, the initial ansatz state is
\begin{equation}
    \label{eq:initial_ansatz_state}
    \ket{\psi^{(0)}}
    =
    S_Y\,S_\Omega\,S_X\ket{0}_X\ket{0}_\Omega\ket{0}_Y
    =
    \frac{1}{\sqrt{|X|}}
    \sum_{x \in X}
    \sum_{\omega \in \Omega}
    \sqrt{\mu(\omega)}\,
    \ket{x}_X\ket{\omega}_\Omega\ket{s_{x,\omega}}_Y.
\end{equation}

\subsection{Ansatz layer}

For a single ansatz layer $k$, let the input state be $\ket{\psi^{(k-1)}} \in \mathcal{H}$. The layer begins with the first-stage phase-separation operator $U_C(\gamma) = e^{-i \gamma \hat C}$, which applies a cost-proportional phase rotation, where $\gamma \in \mathbb R$ is a variational parameter and $\hat C$ is the diagonal operator on the solution register with $\hat C\ket{x} = C(x)\ket{x}$. Applied in layer $k$, this gives $\ket{\psi_C^{(k)}} = U_C(\gamma_k)\ket{\psi^{(k-1)}}$, with amplitudes
\begin{equation*}
    (\psi_C^{(k)})_{x,\omega,y}
    =
    \psi^{(k-1)}_{x,\omega,y} \, e^{-i\gamma_k C(x)}.
\end{equation*}

The recourse oracle computes a cost $Q(x,y;\omega)$ for each encoded computational basis state. The recourse phase-separation operator $U_Q(\gamma)$ then imprints this cost as a phase,
\begin{equation}
\label{eq:U_Q_oracle}
U_Q(\gamma)
\ket{x}_X \ket{\omega}_\Omega \ket{y}_{Y}
=
e^{-i\gamma Q(x,y;\omega)}
\ket{x}_X \ket{\omega}_\Omega \ket{y}_{Y}.
\end{equation}

After the recourse phase in layer $k$, the combined phase-separated state is
\begin{equation}
    \label{eq:phase_separated_state}
    \ket{\psi_{CQ}^{(k)}}
    =
    \sum_{x,\omega,y}
    \psi^{(k-1)}_{x,\omega,y} \,
    e^{-i\gamma_k\!\left[C(x) + Q(x,y;\omega)\right]}
    \ket{x}_X\ket{\omega}_\Omega\ket{y}_{Y}.
\end{equation}
The two phase-separation operators therefore encode the combined first-stage and recourse cost $C(x) + Q(x,y;\omega)$ on each computational basis state.

The ansatz next maps the prepared scenario--recourse component associated with each first-stage decision $x$ to the all-zero state. Define the scenario--recourse reference state as
\begin{equation}
\label{eq:conditional_reference_state}
    \ket{r_x}_{\Omega Y}
    :=
    \sum_{\omega\in\Omega}
    \sqrt{\mu(\omega)}\,
    \ket{\omega}_\Omega\ket{s_{x,\omega}}_Y,
\end{equation}
and the corresponding prepared reference subspace as
\begin{equation*}
    \mathcal S
    :=
    \operatorname{span}
    \left\{
        \ket{x}_X\ket{r_x}_{\Omega Y}:x\in X
    \right\}.
\end{equation*}
Because $S_YS_\Omega$ maps
$\ket{x}_X\ket{0}_\Omega\ket{0}_Y$ to
$\ket{x}_X\ket{r_x}_{\Omega Y}$, applying the inverse preparation unitaries maps
$\mathcal S$ to the flagged subspace $\mathcal H_X\otimes\ket{0}_\Omega\ket{0}_Y$. Acting on the phase-separated state,
this gives
\begin{equation}
\label{eq:flagged_reference_component}
    S_\Omega^\dagger S_Y^\dagger
    \ket{\psi_{CQ}^{(k)}}
    =
    \sum_{x\in X}
    \phi_x^{(k,\mu)}
    \ket{x}_X\ket{0}_\Omega\ket{0}_Y
    +
    \ket{\chi_\perp^{(k)}},
\end{equation}
where $\ket{\chi_\perp^{(k)}}$ has no support on
$\ket{0}_\Omega\ket{0}_Y$. Writing
$a_s(y\mid x,\omega):=\braket{y}{s_{x,\omega}}$, the amplitude of the flagged component associated with first-stage solution $x$ is then
\begin{equation}
\label{eq:flagged_amplitude}
\begin{aligned}
    \phi_x^{(k,\mu)}
    &=
    \bra{x}_X\bra{r_x}_{\Omega Y}
    \ket{\psi_{CQ}^{(k)}} \\
    &=
    e^{-i\gamma_k C(x)}
    \sum_{\omega\in\Omega}
    \sqrt{\mu(\omega)}
    \sum_{y\in Y}
    a_s^*(y\mid x,\omega)\,
    \psi^{(k-1)}_{x,\omega,y}\,
    e^{-i\gamma_k Q(x,y;\omega)}.
\end{aligned}
\end{equation}

The mixing unitary $U_W(t)=e^{-itA_X}$ on the first-stage register is then applied only to this component by conditioning its execution on $\ket{0}_\Omega\ket{0}_Y$. Defining $\Pi_0:=\ket{0,0}\!\bra{0,0}_{\Omega Y}$ on the scenario and recourse registers, the resulting unitary is
\begin{equation}
    \Lambda_0\!\left(U_W(t)\right)
    := U_W(t)\otimes\Pi_0 + \mathbb{I}_X\otimes(\mathbb{I}_{\Omega Y}-\Pi_0).
\end{equation}
The recourse and scenario preparation unitaries are then reapplied so that, altogether, one QASCO layer acts as
\begin{equation}
\label{eq:ansatz_layer_output}
    \ket{\psi^{(k)}}
    =
    S_YS_\Omega\,
    \Lambda_0\!\left(U_W(t_k)\right)
    S_\Omega^\dagger S_Y^\dagger\,
    U_Q(\gamma_k)U_C(\gamma_k)
    \ket{\psi^{(k-1)}}.
\end{equation}

\subsection{Quantum walk interpretation}

\begin{figure}[t]
    \centering
    \includegraphics[width=\linewidth]{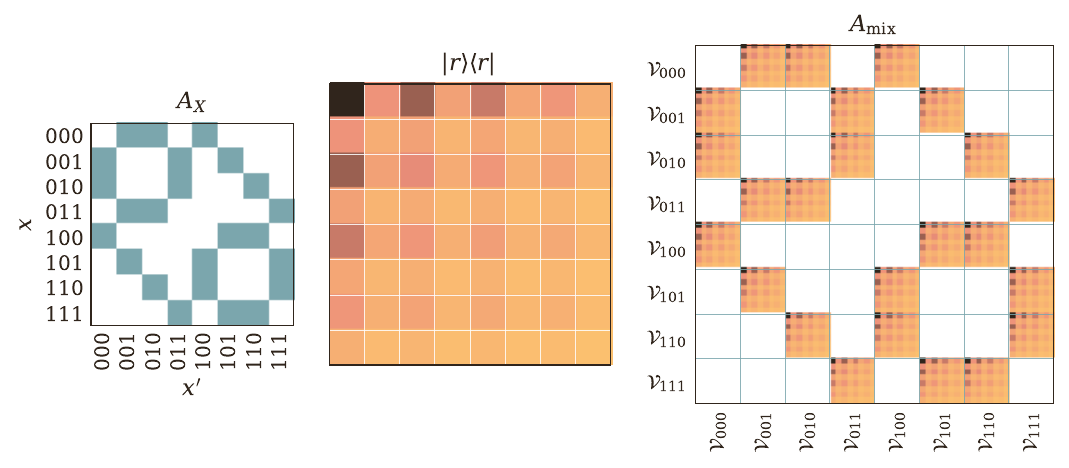}
    \caption{Induced partite-graph structure of the controlled mixer with a transverse-field, or hypercube, solution-space adjacency $A_X$ with $|X|=8$. Here $S_Y$ prepares a uniform superposition over recourse actions, $\ket{s_{x,\omega}}_Y=\ket{s}_Y$. The scenario distribution is a product of $\mu_0=(0.825,0.175)$, $\mu_1=(0.660,0.340)$, and $\mu_2=(0.739,0.261)$. \textbf{Left}: the first-stage solution-space adjacency. \textbf{Middle}: the rank-one projector $\ket{r}\!\bra{r}$, where $\ket{r}=\ket{\mu}_\Omega\ket{s}_Y$, with basis states ordered as $(\omega,y)$. \textbf{Right}: the induced adjacency matrix $A_{\mathrm{mix}}$. Block rows and columns correspond to the partite sets $\mathcal V_x$ and $\mathcal V_{x'}$; each block is $(A_X)_{x,x'}\ket{r}\!\bra{r}$. White entries are zero and darker entries have larger weights.}
    \label{fig:mixer-partite-adjacency}
\end{figure}

To motivate the QASCO mixing mechanism, observe that, since $\Pi_0^2=\Pi_0$, the zero-controlled walk has the exponential form
\begin{equation}
    \label{eq:controlled_mixer_exponential}
    \Lambda_0(U_W(t))
    =
    \exp[-it(A_X\otimes\Pi_0)].
\end{equation}
After conjugation by $S_YS_\Omega$, the induced generator $A_{\mathrm{mix}}=\sum_{x,x'\in X}(A_X)_{x,x'}\ket{x,r_x}\!\bra{x',r_{x'}}$ admits a partite-graph interpretation. For each first-stage decision $x$, we define the partite set
\begin{equation*}
    \mathcal V_x = \{(x, \omega, y) : \omega \in \Omega,\, y \in Y\},
\end{equation*}
with size $|\Omega|\,|Y|$. Each solution-space edge $(A_X)_{x,x'}\neq 0$, with $x\neq x'$, induces connections between all elements with nonzero reference-state amplitudes in the corresponding partite sets $\mathcal V_x$ and $\mathcal V_{x'}$, with weights given by
\begin{equation}
\begin{aligned}
    &\bra{x,\omega,y}A_{\mathrm{mix}}\ket{x',\omega',y'} \\
    &\quad= (A_X)_{x,x'}\braket{\omega,y}{r_x}\braket{r_{x'}}{\omega',y'} \\
    &\quad= (A_X)_{x,x'}\sqrt{\mu(\omega)\mu(\omega')}\,
    a_s(y\mid x,\omega)a_s^*(y'\mid x',\omega').
\end{aligned}
\end{equation}
Thus, amplitude transferred into a first-stage decision is distributed across its scenario--recourse states according to the corresponding reference state. The mixer therefore acts on one aggregated amplitude per first-stage decision, incorporating scenario-weighted recourse phase information through $\phi_x^{(k,\mu)}$ (\cref{eq:flagged_amplitude}). \Cref{fig:mixer-partite-adjacency} illustrates the induced partite-graph structure for the QASCO MaxCut formulation explored later in this work when $S_Y$ is uniform across all recourse choices.

\subsection{Variational objective}
\label{sec:variational_objective}

Altogether, the ansatz prepares the state $\ket{\psi(\boldsymbol\theta)} = \ket{\psi^{(p)}}$,
where $p$ is the number of ansatz layers and $\boldsymbol\theta=(\boldsymbol{\gamma},\boldsymbol{t})$ are $2p$ variational parameters. The variational parameters are optimised to minimise the expectation value of the combined first-stage and recourse cost,
\begin{equation}
\label{eq:ansatz_objective}
f_{\mathrm{joint}}(\boldsymbol\theta)
=\langle\psi(\boldsymbol\theta)|\hat H|\psi(\boldsymbol\theta)\rangle,
\end{equation}
where $\hat H$ is a diagonal cost operator with eigenvalues $C(x)+Q(x,y;\omega)$.  This objective averages the combined first-stage and recourse cost under the measured ansatz distribution, whereas the problem objective $J_\mu$ (see \cref{eq:problem_objective}) combines $C(x)$ with the minimum recourse cost averaged under the target scenario distribution. Minimising $f_{\mathrm{joint}}$ therefore does not by itself
guarantee concentration on decisions that minimise $J_\mu$.

Instead, the problem costs enter via the QASCO mixing process. Expanding the reference state in
\cref{eq:conditional_reference_state} gives
\begin{equation}
\begin{aligned}
F_s^\mu(x)
&:=\langle x,r_x|\hat H|x,r_x\rangle\\
&=\sum_{\omega,y}\mu(\omega)|a_s(y\mid x,\omega)|^2
[C(x)+Q(x,y;\omega)]\\
&=C(x)+\sum_{\omega,y}\mu(\omega)|a_s(y\mid x,\omega)|^2
Q(x,y;\omega),
\end{aligned}
\end{equation}
which we term the \emph{surrogate objective}. It averages the vertex costs within $\mathcal V_x$ according to the reference-state probabilities. For a given $x$, the recourse term averages $Q(x,y;\omega)$ over the target scenario distribution and the prepared recourse states. Since $\hat H$ is diagonal in $x$, the reference component $\sum_x b_x|x,r_x\rangle$ contributes $\sum_x|b_x|^2F_s^\mu(x)$ to the measured objective. While the full ansatz objective also includes the orthogonal component and its interference with the reference component, numerical results in \cref{sec:numerical_results} suggest that changes in the reference contribution account for most of the observed reduction in the measured objective over the problem instances studied in this work.

\subsection{Recourse preconditioning}
\label{sec:recourse_preparation}

The choice of recourse preparation $S_Y$ determines how recourse costs are aggregated and, hence, how closely $F_s^\mu$ approximates $J_\mu$. For any normalised recourse state supported on $Y(x,\omega)$,
\begin{equation*}
\sum_{y\in Y}
|a_s(y\mid x,\omega)|^2 Q(x,y;\omega)
\geq Q^*(x;\omega),
\end{equation*}
hence $F_s^\mu(x)\geq J_\mu(x)$.

The ideal prepared state is supported only on feasible recourse actions minimising $Q(x,y;\omega)$, in which case $F_s^\mu(x) = J_\mu(x)$. However, preparing such a state is generally as hard as solving the recourse problem itself. For example, starting from a uniform recourse state over $Y(x,\omega)$, amplitude amplification requires $O(\sqrt{|Y(x,\omega)|/M})$ oracle calls to amplify the $M$ recourse actions below a chosen cost threshold \cite{BrassardEtAl2002AmplitudeEstimation}. When few actions qualify and $|Y(x,\omega)|\propto|X|$, this can lead to prohibitively deep circuits.

Hence, we consider methods that prepare states merely biased toward low recourse costs within each $(x, \omega)$ block. In this work, we consider a \emph{recourse preconditioner} based on a depth-$p_R$ QWOA circuit with parameters optimised in advance to minimise $\braket{Q}$, averaged over a uniform first-stage distribution and the target scenario distribution $\mu$. We call these parameters the \emph{recourse-preconditioner parameters}, and $\boldsymbol\theta$, which governs the $p$ QASCO layers, the \emph{ansatz parameters}. At $p_R=0$, uniform recourse preparation gives
\begin{equation*}
\ket{s_{x,\omega}}_Y
:=
\frac{1}{\sqrt{|Y(x,\omega)|}}
\sum_{y\in Y(x,\omega)}\ket{y}_Y,
\end{equation*}
for which
\begin{equation*}
F_s^\mu(x)
=
C(x)+\sum_{\omega\in\Omega}\mu(\omega)
\frac{1}{|Y(x,\omega)|}\sum_{y\in Y(x,\omega)}Q(x,y;\omega),
\end{equation*}
Thus, uniform recourse preparation replaces the minimum recourse cost in $J_\mu$ with an average over candidate recourse actions. 

\section{QASCO for stochastic MaxCut}
\label{sec:maxcut_implementation}

We now describe a circuit implementing QASCO (see \cref{sec:ansatz}) for the stochastic MaxCut problem defined in \cref{sec:maxcut}. The solution and recourse registers each require $n$ qubits. The scenario register requires
\begin{equation*}
n_\Omega
=
\sum_{e\in E}
\left\lceil\log_2|\Delta_e|\right\rceil
\end{equation*}
qubits. The configuration-space encoding therefore requires $2n+n_\Omega$ qubits.

We use $b$-bit fixed-point weights and assume each scenario subregister can be reversibly decoded into its numerical edge perturbation using $O(b+\log|E|)$ gates. This holds when each edge has at most a fixed number of possible perturbation values.

For the mixing unitary $U_W(t)$, we use the transverse-field mixer on $X$~\cite{Farhi2014QAOA}. This choice performs well for the weighted MaxCut problem and is compatible with the three-parameter schedule in \cref{eq:linear_ramp}~\cite{bennett2025benchmarkingquantumheuristicsnonvariational}. The mixer requires $n$ single-qubit Pauli-$X$ rotations. In QASCO, these rotations are conditioned on the all-zero state of the scenario and recourse registers. This controlling state can be detected using a Toffoli ladder that computes the logical AND over the $n+n_\Omega$ controls, requiring $n+n_\Omega-2$ work ancillas and one flag qubit, which then controls the $n$ rotations. After decomposition of the Toffoli gates into one- and two-qubit gates, the controlled mixer requires $O(n+n_\Omega)$ two-qubit gates~\cite{BarencoEtAl1995}.

Preparing the initial state on $X$ and the first-stage phase-separation unitary also follows the standard QAOA form~\cite{Farhi2014QAOA,bennett2025benchmarkingquantumheuristicsnonvariational}. We prepare the uniform superposition over $X$ using $n$ Hadamard gates, and $U_C(\gamma)$ decomposes into one two-qubit $ZZ$ rotation for each weighted edge. Using the standard CNOT--$R_Z$--CNOT decomposition, this requires $2|E|$ CNOT gates and $|E|$ single-qubit rotations with no ancillary qubits.

To prepare the scenario reference state, we require unitaries $S_{\Omega_e}$ that prepare the amplitudes $\sqrt{\mu_e(\delta)}$ over $\delta\in\Delta_e$. We can realise each $S_{\Omega_e}$ using standard arbitrary state-preparation methods. For example, the uniformly controlled-rotation method of~\cite{Mottonen2004} requires $O(|\Delta_e|)$ single- and two-qubit gates on $\lceil\log_2|\Delta_e|\rceil$ qubits and no additional ancillas, giving $O(\sum_{e\in E}|\Delta_e|)$ two-qubit gates for $S_\Omega$. 

For the coherent implementation of the piecewise recourse cost, we introduce indicators of whether the original and candidate repaired cuts fall below the threshold,
\begin{equation*}
f_0 = \mathbf 1\{W(x;\omega)<T\}, \qquad
f_1 = \mathbf 1\{W(y;\omega)<T\}.
\end{equation*}
We can then express the recourse cost of \cref{eq:Q_piecewise_maxcut} as
\begin{equation}
\label{eq:Q_compact}
Q(x,y;\omega)
=
f_0(1-f_1)\,d_H(x,y) + f_0f_1\,\Gamma.
\end{equation}

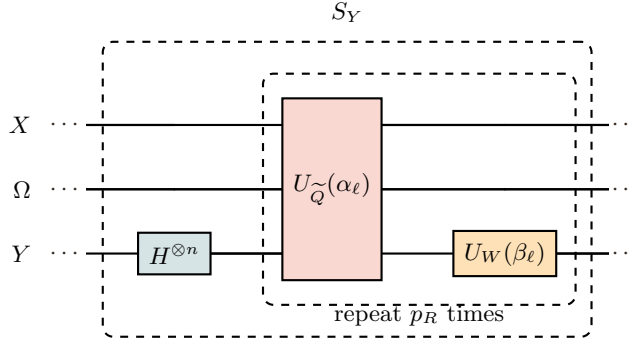
\begin{figure}[t]
    \centering
    \def\recourseoverlay{%
        \foreach \row in {1,2,3}{%
            \node[fill=white,inner sep=1pt,text=pawseyBlack] at (\tikzcdmatrixname-\row-1.center) {$\cdots$};
            \node[fill=white,inner sep=1pt,text=pawseyBlack] at (\tikzcdmatrixname-\row-5.center) {$\cdots$};
        }
    }
    {\fontsize{9}{10.5}\selectfont
    \begin{quantikz}[
        row sep={0.85cm,between origins},
        column sep=0.95cm,
        execute at end picture={\recourseoverlay}
    ]
        \lstick{$X\quad$}\qw
            & \qw
              \gategroup[
                  wires=3,steps=3,
                  style={dashed,rounded corners,inner xsep=10pt,inner ysep=18pt},
                  label style={label position=above,yshift=0.15cm}
              ]{$S_Y$}
            & \gate[3,style={fill=pawseyPeachy!35}]{U_{\widetilde Q}(\alpha_\ell)}
              \gategroup[
                  wires=3,steps=2,
                  style={dashed,rounded corners,inner xsep=4pt,inner ysep=6pt},
                  label style={label position=below,anchor=north,yshift=-0.08cm}
              ]{repeat $p_R$ times}
            & \qw
            & \qw \\
        \lstick{$\Omega\quad$}\qw
            & \qw & \qw & \qw & \qw \\
        \lstick{$Y\quad$}\qw
            & \gate[style={fill=pawseyDustyBlue!30}]{H^{\otimes n}}
            & \qw
            & \gate[style={fill=pawseySunYellow!50}]{U_W(\beta_\ell)}
            & \qw
    \end{quantikz}
    }
    \caption{QWOA recourse preconditioner $S_Y$ for stochastic MaxCut. Each of the $p_R$ layers applies the modified recourse-phase oracle $U_{\widetilde Q}$, followed by transverse-field mixing on $Y$. The preconditioner parameters $(\boldsymbol\alpha,\boldsymbol\beta)$ are optimised first and then held fixed during optimisation of the main ansatz parameters. For $p_R=0$, the circuit reduces to $H^{\otimes n}$ on the recourse register.}
    \label{fig:recourse-preconditioner-circuit}
\end{figure}

The recourse preparation $S_Y$ starts with $n$ Hadamard gates, followed by $p_R$ layers alternating phase separation under
\begin{equation}
\label{eq:preconditioner_recourse_cost}
\widetilde Q(x,y;\omega)
=(1-f_1)\,d_H(x,y)+f_1\,\Gamma
\end{equation}
with transverse-field mixing on $Y$ (as illustrated in \cref{fig:recourse-preconditioner-circuit}). We denote this phase-separation unitary by $U_{\widetilde Q}$.

To compute $f_0$ and $f_1$, the realised cut values $W(x;\omega)$ and $W(y;\omega)$ are accumulated and compared with the cut threshold value $T$. The cut-value accumulator requires $b+\lceil\log_2|E|\rceil$ qubits and can be reused to compute the two failure flags sequentially. Evaluating the cut values with standard reversible adders requires $O(|E|(b+\log|E|))$ two-qubit gates~\cite{CuccaroDraperEtAl2004}. Including the failure flags and arithmetic workspace, this requires $O(b+\log|E|)$ reusable ancillary qubits. 

The Hamming distance is $d_H(x,y)=\sum_{j=0}^{n-1}(x_j\oplus y_j)$,
for which the first term of \cref{eq:Q_compact} can be phase encoded using $n$ commuting rotations conditioned on $f_0=1$, $f_1=0$, and $x_j\neq y_j$. The penalty term requires a single phase rotation conditioned on $f_0=f_1=1$. After applying these phases, the cut-value calculations and failure flags are uncomputed. After decomposition into one- and two-qubit gates, the recourse phase oracle therefore requires $O(|E|(b+\log|E|)+n)$ two-qubit gates and $O(b+\log|E|)$ reusable ancillary qubits per application. Within $S_Y$, $U_{\widetilde Q}$ computes and uncomputes only $W(y;\omega)$. The depth-$p_R$ recourse preparation therefore requires $O(p_R[|E|(b+\log|E|)+n])$ two-qubit gates per $S_Y$ application, in addition to the single-qubit gates used for the initial and transverse-field mixing layers.

Combining these components, the circuit requires
$2n+n_\Omega+O(\max\{n+n_\Omega,b+\log|E|\})$ qubits, where the final term accounts for reusable work ancillas required by the controlled mixer and recourse phase oracle. The one-time preparation of the initial state requires $O(\sum_{e\in E}|\Delta_e|+p_R[|E|(b+\log|E|)+n])$ two-qubit gates, in addition to $O(n(1+p_R))$ single-qubit gates. Each QASCO layer then applies $U_C$ and $U_Q$, followed by the adjoint scenario and recourse preparation unitaries, the controlled mixer, and the corresponding preparation unitaries. The resulting two-qubit gate complexity per layer is
\[
O\!\left(
(1+p_R)\left[|E|(b+\log|E|)+n\right]
+n_\Omega
+\sum_{e\in E}|\Delta_e|
\right).
\]
With at most a fixed number of possible perturbation values per edge, both $n_\Omega$ and $\sum_e|\Delta_e|$ are $O(|E|)$. The per-layer gate complexity then reduces to $O((1+p_R)[|E|(b+\log|E|)+n])$. For fixed $p_R$, the leading two-qubit gate cost is therefore the repeated implementation of the recourse phase oracle.

Since $|X|=2^n$, this is $O(|E|(b+\log|E|)+\log|X|)$. If, in addition, each edge has at least two possible perturbation values, then $|E|=\Theta(\log|\Omega|)$ and this becomes
\[
O\!\left(
\log|\Omega|\,[b+\log\log|\Omega|]
+\log|X|
\right).
\]
Thus, for fixed $p_R$ and fixed weight precision $b$, the per-layer two-qubit gate complexity is polylogarithmic in $|X|$ and $|\Omega|$, although the joint configuration space has size $|X|^2|\Omega|$.

\section{Numerical methods}
\label{sec:methods}

We conduct a numerical study in two parts. First, we examine QASCO's convergence behaviour and compare it with an adaptation of the QAOA-based method in~\cite{kuroiwa2025}, which we refer to as the \emph{two-stage QAOA}, under full variational optimisation. The two-stage QAOA acts on the same registers $X$, $\Omega$, and $Y$, and begins from the product state $\ket{s}_X\ket{\mu}_\Omega\ket{s}_Y$. In place of the recourse preconditioning and controlled mixing used by QASCO, it applies two sequential QAOA stages with transverse-field mixing. The first stage has $p_1$ layers acting on $X$, with phase separation under the first-stage cost $C$, while the second has $p_2$ layers with phase separation under the recourse cost $Q(x,y;\omega)$ and mixing over $Y$. Following~\cite{kuroiwa2025}, we set $p_1=p_2=p$. QASCO and the two-stage QAOA are optimised using the same ansatz objective function.

The second part of the numerical study evaluates QASCO as a fixed-schedule first-stage candidate generator against classical uniform enumeration of the solution space, random-restart greedy search, and tabu search. For each considered problem setting, we first calibrate the QASCO and configurable classical-search parameters on a calibration set and then evaluate performance on a separate test set. The stochastic objective is approximated using sample average approximation, and the methods are compared by the relative number of cut evaluations, or query cost, required to reach the stochastic optimum. This tests whether a schedule calibrated on representative instances can generate useful first-stage candidates without per-instance variational optimisation.

\subsection{Benchmark ensembles}
\label{sec:benchmark_ensembles}

We performed all experiments using benchmark ensembles of randomly generated stochastic MaxCut problem instances. Graphs are sampled uniformly from connected simple graphs with $n$ labelled vertices and $|E|$ edges. Each edge has an independent $m$-point weight distribution, giving $m^{|E|}$ scenarios. Given generation parameters $a<b$ and a fragile-edge fraction $\rho$, we select $\max\{1,\operatorname{round}(\rho|E|)\}$ edges uniformly as \emph{fragile}. Their weights comprise $m-1$ independent values sampled uniformly from $[a,a+0.05(b-a)]$ and one sampled uniformly from $[3b,8b]$. We sample the probability assigned to the high weight uniformly from $[0.05,0.15]$; the remaining probability is divided among the low weights in proportion to independent values sampled uniformly from $[0,1]$. For each remaining edge, a centre $c$ is drawn uniformly from $[a+0.3(b-a),b-0.3(b-a)]$, with $m$ weights drawn uniformly from $[c-0.05(b-a),c+0.05(b-a)]$ and probabilities obtained by normalising independent values sampled uniformly from $[0,1]$.

Nominal edge weights are the probability-weighted means. Rare high weights on fragile edges can raise these means above the weights typically realised, encouraging disjoint nominal and stochastic optima. We accept only instances with unique, distinct nominal and stochastic optima in $X_{\mathrm{can}}$.

The ansatz-convergence comparison (see \cref{sec:two_stage_qaoa_methods}) between QASCO and the two-stage QAOA uses three ensembles at $n=6$, $m=3$, and $|E|=6$, generated with $\rho=0.6$, $(a,b)=(1,3)$, and recourse threshold $T=T_{\mathrm{med}}$, where $T_{\mathrm{med}}$ is the median expected cut value over the solution space, chosen as it was found to trigger recourse for a substantial fraction of solutions. To explore the extent to which the two ans\"atze can distinguish the nominal and stochastic optima, each ensemble consists of instances with the same number of canonical solutions ranked strictly between the stochastic optimum $X_\mu^*$ and the nominal optimum $X_C^*$ under $J_\mu$, which we denote as the \emph{rank gap}. The three ensembles have $r_g=1,2,3$ with $50$ instances each.

The fixed-schedule calibration and test sets (see \cref{sec:solver_comparison_protocol}) contain instances with $n=6,7,8$ at fixed $m=3$ and $|E|=10$; $m=2,\ldots,5$ at fixed $n=7$ and $|E|=8$; and $|E|=8,9,10$ at fixed $n=7$ and $m=3$, corresponding to a total of eight unique problem settings. Instances are generated with $\rho=0.2$, $(a,b)=(0.2,5.0)$, and recourse threshold $T$ given by the 85th percentile of the expected cut values over the per-instance solution space.

Each set contains equal numbers of instances in repair-depth bins $D\in\{0,1,2\}$. For the stochastic optimum $x\in X_\mu^*$, $D$ is the largest value of $Q^*(x;\omega)$ over scenarios admitting a successful repair, with $D=0$ if no repairable scenario requires a vertex move. For each problem setting $(n,m,|E|)$, the calibration and test sets each contain $60$ accepted instances, with $20$ instances in each repair-depth bin, giving $480$ calibration and $480$ test instances across the eight settings.

\subsection{State-vector simulation}
\label{sec:state_vector_simulation}

We use a GPU-accelerated, distributed-memory exact state-vector simulator that exploits the block structure of the phase operators and the reference-subspace structure of the QASCO mixer. The state is represented in the basis $(x,\omega,y)$, with $|X||\Omega||Y|=2^{2n}m^{|E|}$ amplitudes when every edge has $m$ perturbation values. The largest simulated state space, at $n=7$, $|E|=8$, and $m=5$, contains $2^{14}\cdot5^8$ complex amplitudes and requires approximately $95$~GiB to store in double precision. Simulations were run on the NVIDIA GH200 nodes of Setonix at the Pawsey Supercomputing Research Centre.

During recourse-preconditioner optimisation, scenarios that induce equivalent recourse problems are grouped and represented by their combined probability weights. This preserves expectation values and normalisation over the original scenario distribution. For each QASCO mixing step, the simulator projects the joint state onto the prepared scenario--recourse reference states. Only the resulting $|X|$-dimensional vector undergoes transverse-field mixing; its change is expanded in the same reference states and added to the joint state. This implements the controlled mixer without constructing its full joint-space matrix.

For the two-stage QAOA, the first stage acts only on $X$, so we first evolve a $|X|$-dimensional state vector and then form $\ket{\psi_X}\ket{\mu}_\Omega\ket{s}_Y$. The second stage acts independently within each $(x,\omega)$ sector on the recourse register. The objective and marginal probabilities are obtained by summing the contributions from these sectors.

\subsection{Ansatz convergence}
\label{sec:two_stage_qaoa_methods}

We investigate the ability of the QASCO mixing unitary to drive convergence of the marginal distribution over the $X$ register towards the stochastic optimum, such that solutions in $X_\mu^*$ are preferentially amplified. We assess this across the three $(n=6, |E|=6, m=3)$ rank-gap ensembles described in \cref{sec:benchmark_ensembles}, comparing the convergence of QASCO with that of the two-stage QAOA.

For QASCO, we set the recourse-preconditioner depth $p_R=p$. We use recourse-first optimisation for both ans\"atze: the $2p$ recourse-related parameters are first optimised to minimise the expected measured recourse cost $\langle Q\rangle$, with a uniform first-stage distribution and the target scenario distribution $\mu$. These parameters are then held fixed while the remaining $2p$ parameters are optimised to minimise $f_{\mathrm{joint}}$. For QASCO, these are the ansatz parameters; for the two-stage QAOA, they are the first-stage parameters. Both methods therefore optimise $4p$ parameters per instance and depth. The order of circuit operations is unchanged.

Optimisation is performed using the Nelder--Mead simplex method~\cite{nelder1965simplex}. Each optimisation uses ten independent starting points at $p=1$ and retains the best result. At each greater depth, one attempt is warm-started from the preceding depth's best parameters using the INTERP rule of~\cite{zhou_quantum_2020}. We also test joint optimisation of all $4p$ two-stage QAOA parameters under $f_{\mathrm{joint}}$, as originally formulated in~\cite{kuroiwa2025}, using the same depth-one restart count and INTERP initialisation. All protocols allow a maximum of $5000$ iterations per optimisation attempt. For consistency with QASCO, the main comparison uses recourse-first optimisation.

Convergence is assessed from the marginal probabilities of the canonical first-stage cuts. For a canonical cut $x$, we denote by $P_p(x)$ the sum of the probabilities of measuring $x$ and its bitwise complement at ansatz depth $p$. We first report rank-resolved probability by ordering the canonical cuts by increasing stochastic objective $J_\mu$, so rank one corresponds to $X_\mu^*$, and examine how probability redistributes across the best-ranked cuts as ansatz depth increases. We then compare the probability assigned to the stochastic optimum $X_\mu^*$ with that assigned to the nominal optimum $X_C^*$. For each instance and depth, we first average probabilities over the optimisation repeats. We report the median of these per-instance probabilities across the 50 instances in each rank-gap ensemble, while the comparison between $X_\mu^*$ and $X_C^*$ uses the averaged probabilities for each individual instance.

\subsection{Candidate-generator benchmark}
\label{sec:solver_comparison_protocol}
\label{sec:shared_recourse_evaluator}

We compare fixed-schedule QASCO with uniform enumeration, random-restart greedy search, and fixed-tenure random-restart tabu search as first-stage candidate generators. For each problem setting, we configure the methods using the fixed-schedule calibration set described in \cref{sec:benchmark_ensembles} and then evaluate them on the corresponding test set.

Candidate costs are estimated using sample average approximation (SAA)~\cite{KleywegtShapiroHomemdeMello2002SAA}. Each run uses a fixed batch $\mathcal B=(\omega^{(1)},\ldots,\omega^{(M_0)})$ of $M_0$ scenarios sampled with replacement from $\mu$. For a candidate $x$, we use
\begin{equation}
    \widehat J_{\mathcal B}(x)
    =
    C(x)
    +
    \frac{1}{M_0}
    \sum_{\ell=1}^{M_0} Q^*(x;\omega^{(\ell)}).
    \label{eq:saa_estimator}
\end{equation}
Each $Q^*(x;\omega)$ is evaluated exactly by searching the recourse actions in increasing Hamming distance from $x$ until a successful repair is found or all recourse actions have been checked, with each cut evaluation counted as one query. We calibrate the SAA batch size $M_0$ independently for each problem setting by considering $M=1,2,4,\ldots$ up to the full scenario-space size $m^{|E|}$. For each calibration instance and tested $M$, we use $200$ independent SAA batches. For each instance, we determine the smallest $M$ for which SAA selects a member of $X_\mu^*$ as the minimiser in at least $90\%$ of the $200$ batches, and maintains this success rate at all larger tested batch sizes. We denote this required batch size by $M_{\mathrm{req}}$. Within each of the three repair-depth bins, we take the $90$th percentile of these required batch sizes and set $ M_0$ to the largest of the three. The query budget $B_{\mathrm q}$ for each problem setting is then set to the median query cost of evaluating every candidate in $X_{\mathrm{can}}$ once using the SAA evaluator with batch size $M_0$. Within a run of a candidate generator, we cache the SAA-evaluated cost for each observed candidate, so that repeated evaluations of the same candidate do not incur an additional query cost. Altogether, this design reduces the extent to which differences between candidate generators are confounded by stochastic variation in SAA accuracy or by setting-dependent candidate-evaluation cost.

We record when a stochastic-optimal candidate first completes its SAA evaluation. We compare candidate-generation performance using the expected running time (ERT)~\cite{HansenAugerEtAl2021COCO}. For run $a$, let $N_a^{\mathrm{hit}}$ denote the query count at which a member of $X_\mu^*$ first completes its SAA evaluation, and define the capped query cost $c_a=\min(N_a^{\mathrm{hit}},B_{\mathrm q})$, such that an unsuccessful run incurs the full query budget. The ERT is then
\begin{equation}
\label{eq:ert}
    \mathrm{ERT}
    =
    \frac{\sum_{a=1}^{A} c_a}
    {\#\{a:N_a^{\mathrm{hit}}\le B_{\mathrm q}\}}.
\end{equation}
Here $A$ is the number of runs. We use ERT when calibrating the candidate generators. To compare performance across problem settings with different query budgets, we report the hit rate $h$, the fraction of runs that fully evaluate a stochastic optimum, together with normalised ERT, $\mathrm{ERT}_{\mathrm{norm}}$, where normalised ERT is ERT divided by the setting's query budget $B_{\mathrm q}$.

For each QASCO configuration $(p,p_R)$ and problem setting, we treat each calibration instance independently, first optimising up to three linear-ramp parameters for the recourse preconditioner when $p_R>0$ and then three for the ansatz. The phase and walk times follow the linear-ramp parameterisation of \cref{eq:linear_ramp} throughout optimisation. We average the generated phase and walk schedules across calibration instances, then fit three common parameters for each schedule to this mean by unweighted least squares. These fitted parameters are held fixed when generating candidates on the calibration and test sets. With this schedule, first-stage candidates are generated by repeatedly sampling the $X$ register of the prepared ansatz state, with each circuit shot producing one candidate that is mapped to its canonical representative $x$. For benchmark scoring, we record queries up to the first fully evaluated member of $X_\mu^*$ or exhaustion of $B_{\mathrm q}$. Each depth-$p$ circuit shot with recourse-preconditioner depth $p_R$ is charged
\begin{equation}
N_{\mathrm q}^{\mathrm{shot}}(p,p_R)
=
5p+2p_R+4pp_R,
\label{eq:quantum_shot_charge}
\end{equation}
We count one query for $U_C$ and one for each realised-cut computation or uncomputation. Thus, $5p$ counts $U_C$ and $U_Q$ in the main layers, $2p_R$ counts the initial $S_Y$, and $4pp_R$ counts $S_Y$ and $S_Y^\dagger$ within those layers. Each preconditioner layer uses two queries. SAA candidate evaluation is charged separately. We test $p\in\{1,\ldots,8\}$ and $p_R\in\{0,1,2,4,8\}$ on the calibration set, selecting the configuration that minimises the ERT.

All methods use the same query budget $B_{\mathrm q}$ for each problem setting. Uniform enumeration evaluates the canonical solution space in random order without replacement. Greedy search selects the best strictly improving one-bit neighbour and restarts when no improvement is possible. Tabu search selects the best admissible one-bit neighbour, permitting worsening moves, and prohibits moves involving recently flipped bits for a fixed tenure $\tau$, unless flipping them improves the best solution found so far~\cite{Glover1990TabuSearchPartII,NathKuhnle2025MaxCutBench}. Both searches initialise each restart uniformly among unevaluated canonical solutions. We calibrate $\tau\in\{1,\ldots,n-2\}$ for each problem setting, choosing the tenure that minimises calibration ERT.

\section{Results}
\label{sec:numerical_results}

We first compare QASCO and the two-stage QAOA on the three $n=6$ rank-gap ensembles, each containing $50$ instances, using recourse-first optimisation for both methods as described in \cref{sec:two_stage_qaoa_methods}. We average objective values and first-stage probabilities over optimisation repeats within each instance and depth, and report the median over instances with interquartile ranges where shown. We then compare QASCO with SAA-driven classical candidate generators on $480$ test instances across eight problem settings. Each setting has $60$ test instances and five runs per instance, giving $300$ test runs per method per setting and $2400$ per method overall. For hit rates, ERT and ERT ratios, we report $95\%$ percentile bootstrap confidence intervals obtained by resampling whole instances within repair-depth bins, using the same resampled instances for all methods~\cite{EfronTibshirani1993Bootstrap}.

\subsection{Comparison with two-stage QAOA}
\label{sec:comparison-with-two-stage_qaoa}

\begin{figure}[t]
    \centering
    \includegraphics[width=\linewidth]{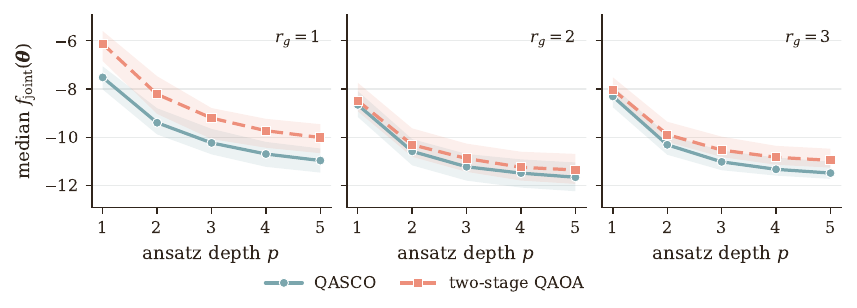}
    \caption{Median optimised joint expectation value $f_{\mathrm{joint}}(\boldsymbol\theta)$ versus ansatz depth $p$ for QASCO and the two-stage QAOA.}
    \label{fig:n6-ansatz-objective-by-rg}
\end{figure}

\begin{figure}[p]
    \centering
    \includegraphics[width=0.85\linewidth]{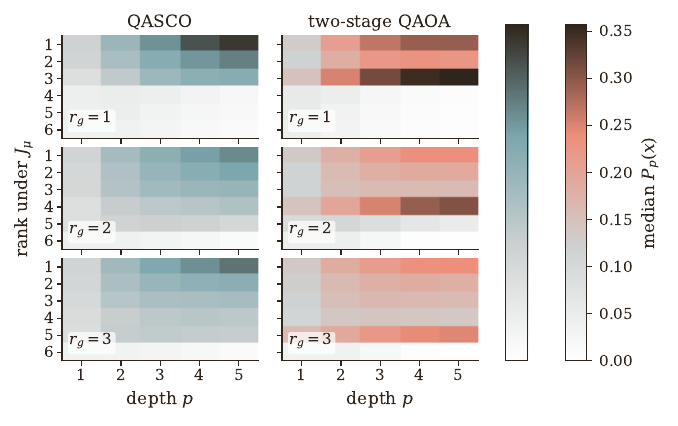}
    \caption{Median first-stage marginal probability $P_p(x)$ by rank under the stochastic objective $J_\mu$ and ansatz depth $p$, where rank one corresponds to the stochastic optimum. Rows are ordered by increasing rank gap $r_g$, with the problem instances in the first row having one solution between the stochastic optimum $X_\mu^*$ and the nominal optimum $X_C^*$, and the second and third rows having two and three such solutions, respectively.}
    \label{fig:n6-ansatz-rank-probability-by-rg}
\end{figure}

\begin{figure}[p]
    \centering
    \includegraphics[width=\linewidth]{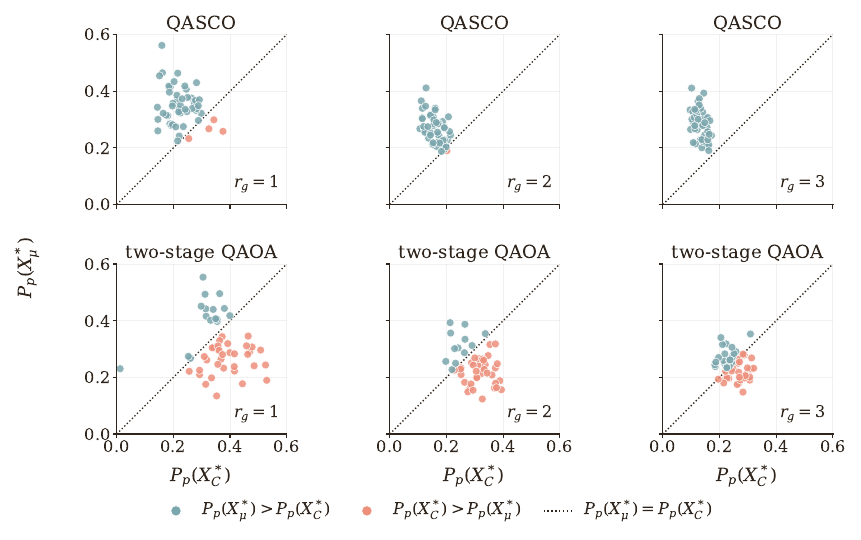}
    \caption{Probability of measuring the stochastic optimum $X_\mu^*$ versus the nominal optimum $X_C^*$ at $p=5$ for each of the rank-gap ensembles. Each point is one problem instance.  Points above the dotted diagonal assign more probability to the stochastic optimum than to the nominal optimum.}
    \label{fig:n6-ansatz-surrogate-first-stage-preference}
\end{figure}

\Cref{fig:n6-ansatz-objective-by-rg} shows that the median optimised joint expectation value $f_{\mathrm{joint}}$ decreases at each simulated depth for both ans\"atze in all three ensembles, with QASCO consistently achieving lower values by a small margin. At $p=5$, the QASCO medians are $-10.961$, $-11.654$, and $-11.481$ for $r_g=1,2,3$, respectively, compared with $-10.008$, $-11.353$, and $-10.955$ for the two-stage QAOA.

In \cref{fig:n6-ansatz-rank-probability-by-rg}, QASCO assigns the largest median probability among the six displayed ranks to the stochastic optimum at every tested depth in all three ensembles. Its median stochastic-optimum probability increases at each depth, from $0.119$, $0.113$, and $0.114$ at $p=1$ to $0.338$, $0.263$, and $0.283$ at $p=5$ for $r_g=1,2,3$, respectively. The two-stage QAOA starts with slightly higher median stochastic-optimum probabilities of $0.131$, $0.135$, and $0.134$, which reach $0.293$, $0.235$, and $0.237$ at $p=5$. However, its largest median probability remains at the nominal optimum at all considered depths.

\Cref{fig:n6-ansatz-surrogate-first-stage-preference} compares the stochastic- and nominal-optimum probabilities within each instance at $p=5$. QASCO assigns more probability to $X_\mu^*$ than to $X_C^*$ in $46/50$, $49/50$, and $50/50$ instances for $r_g=1,2,3$, respectively. The corresponding counts for the two-stage QAOA are $16/50$, $12/50$, and $21/50$. The QASCO counts increase across these ensembles, while the two-stage QAOA counts are nonmonotonic in the rank gap.

In \cref{sec:appendix_objective_contributions}, we further examine the QASCO state in the reference subspace $\mathcal S$ and its orthogonal complement, together with their contributions to the change in $f_{\mathrm{joint}}$. Across the three rank-gap ensembles, the median reference-subspace occupation $P_R$ is $0.765$ at $p=1$ and $0.849$ at $p=5$, compared with $P_\perp=0.235$ and $0.151$, respectively. At $p=5$, the median cumulative changes are $\Delta R_5=-5.881$ and $\Delta I_5=-0.434$. Overall, the state remains primarily in $\mathcal S$ at every tested depth, with changes in the reference contribution accounting for most of the observed reduction in $f_{\mathrm{joint}}$. This is consistent with a substantial role for the surrogate objective $F_s^\mu$ in the measured convergence.

With recourse-first optimisation, the cumulative evaluation counts are similar for the two ans\"atze. The corresponding ensemble medians are $5310$--$5418.5$ for QASCO and $5549$--$5835$ for the two-stage QAOA. The classical optimiser reports successful termination for all instances. We found that joint optimisation of the two-stage QAOA variational parameters yields final median joint expectation values of $-9.014$, $-10.382$, and $-9.710$ for $r_g=1,2,3$, respectively, all higher than those obtained with recourse-first optimisation. The corresponding median stochastic-optimum probabilities are $0.2947$, $0.2354$, and $0.2373$, differing from the recourse-first values by less than $0.002$ in each ensemble. However, recourse-first optimisation reduces cumulative objective-function evaluations through $p=5$ by about $60\%$ compared with the joint-optimisation ensemble medians of $13728$--$13948.5$. 

\subsection{Candidate-generator benchmark}
\label{sec:solver_benchmark}

This fixed-schedule benchmark compares QASCO with classical uniform enumeration, greedy search, and tabu search on the eight $(n,m,|E|)$ problem settings described in \cref{sec:benchmark_ensembles}. QASCO uses a calibration-selected $(p,p_R)$ configuration for each setting, and tabu search uses a calibration-selected tenure $\tau$.

\begin{figure}[t]
    \centering
    \includegraphics[width=\linewidth]{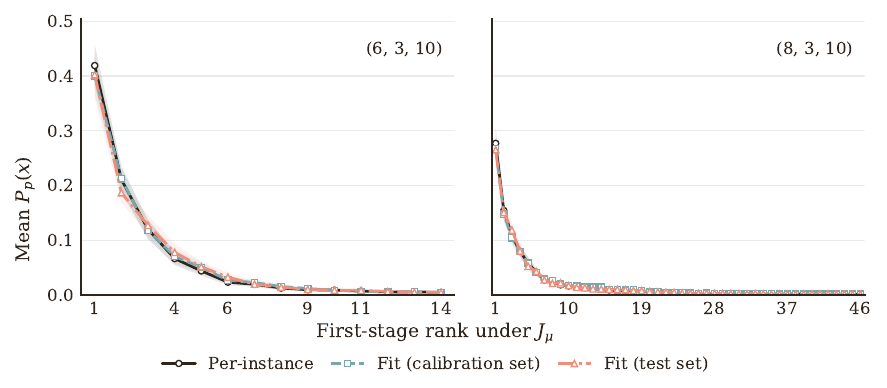}
    \caption{Mean QASCO first-stage marginal probability by rank under the stochastic objective $J_\mu$ for \textbf{(left)} $n=6$, $m=3$, $|E|=10$, $(p,p_R)=(8,0)$ and \textbf{(right)} $n=8$, $m=3$, $|E|=10$, $(p,p_R)=(7,1)$. Curves compare per-instance optimisation of the recourse-preconditioner and ansatz parameters on the calibration set with the calibrated schedule applied to the calibration and test sets. Each panel shows a range of ranks containing at least $95\%$ of each mean distribution without renormalisation.} 
    \label{fig:fixed-schedule-transfer}
\end{figure}

In \cref{fig:fixed-schedule-transfer}, we show representative probability distributions for the $n=6$ and $n=8$ problem settings with $m=3$ and $|E|=10$. In both cases, the calibrated schedule produces similar distributions on the calibration and test sets. For $n=6$, mean stochastic-optimum probabilities are $0.419$ with per-instance optimisation on the calibration set, $0.400$ with the calibrated schedule on calibration instances, and $0.402$ with the calibrated schedule on test instances. For $n=8$, the corresponding probabilities are $0.277$, $0.267$, and $0.265$. Across the three distributions, the eight lowest-cost cuts account for a mean total probability of $0.911$--$0.915$ for $n=6$ and $0.751$--$0.769$ for $n=8$. Across all eight settings, the test-minus-calibration difference in the ensemble mean of the stochastic objective $J_\mu$, using the same calibrated schedule on both sets, ranges from $-0.327$ to $+0.587$, with all $95\%$ confidence intervals containing zero. 

\begin{figure}[p]
    \centering
    \includegraphics[width=\linewidth]{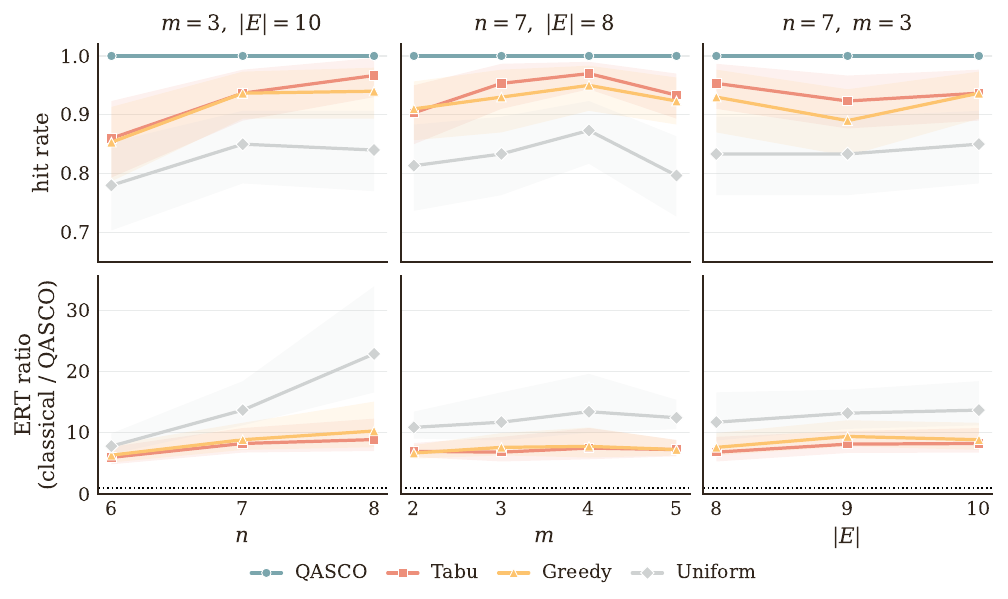}
    \caption{Hit rate $h$ (top row) and classical-to-QASCO ERT ratio (bottom row) for the candidate-generator benchmark. Columns vary $n$ at fixed $m=3$, $|E|=10$; $m$ at fixed $n=7$, $|E|=8$; and $|E|$ at fixed $n=7$, $m=3$. ERT ratios greater than one favour QASCO, and the dotted line at one denotes equal ERT.}
    \label{fig:solver-scaling-performance}
\end{figure}

\Cref{fig:solver-scaling-performance} shows that QASCO returns a stochastic optimum in all $2400$ runs, with hit rate $1.000$ in every problem setting. Classical hit rates range from $0.860$ to $0.970$ for tabu search, $0.853$ to $0.950$ for greedy search, and $0.780$ to $0.873$ for uniform enumeration. QASCO has the lowest ERT in every setting.   \Cref{sec:appendix_full_solver_comparison_table} reports ERT values, query budgets, SAA batch sizes, and calibrated tabu tenures; \cref{sec:appendix_linear_ramp_schedule_parameters} reports the QASCO depths and schedule parameters.

At fixed $m=3$ and $|E|=10$, the ERT ratios increase in favour of QASCO at each tested $n$, from $6.0$ to $8.9$ for tabu, $6.3$ to $10.3$ for greedy, and $7.8$ to $22.9$ for uniform enumeration between $n=6$ and $n=8$. Varying $m$ with $n=7$ and $|E|=8$, all three ratios are largest at $m=4$ and decrease at $m=5$. With varying $|E|$ at fixed $n=7$ and $m=3$, the tabu and uniform ratios increase from $6.8$ to $8.2$ and $11.7$ to $13.7$, respectively, while the greedy ratio increases from $7.6$ at $|E|=8$ to $9.4$ at $|E|=9$, then decreases to $8.9$ at $|E|=10$.

\begin{figure}[p]
    \centering
    \includegraphics[width=\linewidth]{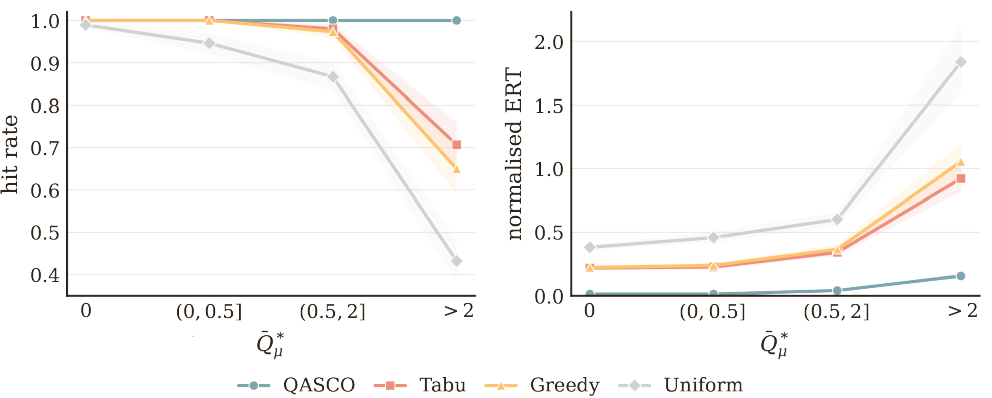}
    \caption{Candidate-generator performance grouped by expected minimum recourse cost at the stochastic optimum, $\bar Q_\mu^*:=\mathbb E_{\omega\sim\mu}[Q^*(x;\omega)]$ for $x\in X_\mu^*$, pooled over the eight problem settings. The groups $\bar Q_\mu^*=0$, $0<\bar Q_\mu^*\le0.5$, $0.5<\bar Q_\mu^*\le2$, and $\bar Q_\mu^*>2$ contain $151$, $104$, $120$, and $105$ test instances, respectively. \textbf{Left:} Hit rate. \textbf{Right:} Normalised ERT.}
    \label{fig:solver-recourse-conditioned}
\end{figure}

\Cref{fig:solver-recourse-conditioned} groups the results by expected minimum recourse cost at the stochastic optimum. QASCO has a hit rate of $1.000$ in all four groups. Tabu and greedy search also have hit rate $1.000$ in the two lowest-cost groups, while uniform enumeration has hit rates $0.989$ and $0.946$. In the highest-cost group, $\bar Q_\mu^*>2$, hit rates are $0.707$ for tabu, $0.650$ for greedy, and $0.432$ for uniform enumeration. Normalised ERT increases across the four groups for all methods. Between the zero- and highest-cost groups, it increases from $0.013$ to $0.156$ for QASCO, $0.219$ to $0.922$ for tabu, $0.223$ to $1.055$ for greedy, and $0.382$ to $1.837$ for uniform enumeration. QASCO has the lowest normalised ERT in every group. We examine the query complexity of classical recourse evaluation and the SAA sampling requirements in \cref{sec:appendix_evaluation_sampling_costs}.

\section{Discussion}
\label{sec:discussion}

The numerical results show that QASCO distinguishes stochastic-optimal first-stage decisions from the nominal optimum. In \cref{fig:n6-ansatz-rank-probability-by-rg} we see that, with variational optimisation of all $2p$ parameters, QASCO increasingly concentrates probability on the stochastic optimum under $J_\mu$ as depth increases and generally assigns it more probability than the nominal optimum. In contrast, the two-stage QAOA increasingly concentrates on the nominal optimum.

The observed preference of the two-stage QAOA for the nominal optimum may reflect the separation of its subcircuits over the first and second stages. The first-stage QAOA circuit redistributes probability using phase separation under $C(x)$, while the subsequent recourse QAOA circuit preserves the first-stage marginal~\cite{kuroiwa2025}. Although recourse costs can still influence the optimised first-stage parameters through $f_{\mathrm{joint}}$, they do not enter the first-stage phase separator, which may weaken the link between the ansatz dynamics and the stochastic optimum. In contrast, the QASCO mixing unitary aggregates amplitudes with recourse-dependent phases, so this information more directly influences the marginal distribution of the first-stage register (\cref{eq:flagged_amplitude}). The extra degrees of freedom enabled by the separate QAOA subcircuits of the two-stage QAOA may allow superior amplification of the stochastic optimum with appropriate parameters, but optimising the joint objective $f_\mathrm{joint}$ alone may make such parameters difficult to identify. With recourse-first optimisation, both ans\"atze optimise $4p$ parameters in two steps and were observed to converge consistently during classical parameter optimisation, with similar numbers of objective-function evaluations. QASCO's stronger preference for the stochastic optimum therefore persists under comparable parameter optimisation effort.

Parameter optimisation overhead and satisfying problem constraints are significant challenges for quantum variational algorithms in general~\cite{FuchsEtAl2022ConstraintMixers,Cerezo2021VQA}. To this end, the modular design of the QASCO circuit lets existing QAOA and QWOA state-preparation and mixing unitaries serve as $S_X$ and $U_W$, respectively, and, where appropriate, as the recourse preparation $S_Y$. In \cref{sec:solver_benchmark}, we show that this enables a linear-ramp parameterisation scheme developed for deterministic optimisation in the QWOA framework \cite{BennettNoakesWang2024NonVariational,BennettNoakesWang2024Analysis}. This uses three classically optimised ansatz parameters and at most three for the recourse preconditioner, requiring only the ansatz parameters when $p_R=0$.

The distinction between the efficiently implementable joint objective $f_{\mathrm{joint}}$, the stochastic objective $J_\mu$ and the surrogate objective $F_s^\mu$ remains an important consideration. Our numerical results show useful first-stage selection when optimising $f_{\mathrm{joint}}$, despite the differences between these quantities. Three of the eight calibration-selected configurations use $p_R=0$ (see \cref{sec:appendix_linear_ramp_schedule_parameters}), indicating that recourse preconditioning is not always necessary for effective candidate generation when stochastic-optimal decisions remain favourable under the surrogate objective. Whether alternative recourse preconditioning strategies, such as threshold-based amplitude amplification, can improve convergence or yield a more efficient circuit realisation is an area for future investigation.

With the linear-ramp parameterisation, QASCO can prepare useful first-stage distributions without per-instance parameter optimisation on the test set. In the two settings illustrated in \cref{fig:fixed-schedule-transfer}, calibrated schedules retain much of the stochastic-optimum probability and low-cost concentration achieved by per-instance optimisation on the calibration set, and produce similar cost-ranked distributions on calibration and test instances. These observations suggest that we may amortise parameter-optimisation overhead by calibrating schedule parameters offline on representative problem instances.

Under the stated query-cost model, QASCO with calibrated parameters attains the lowest ERT in every tested setting and reaches a stochastic optimum in every run. In \cref{fig:solver-scaling-performance}, the classical-to-QASCO ERT ratio increases with the size of the first-stage space, consistent with the increasing advantage reported for deterministic MaxCut over similar simple classical heuristics~\cite{bennett2025benchmarkingquantumheuristicsnonvariational}. Classical hit rates also generally improve across this range; one possible explanation is that, as $|E|$ is held constant, the graph becomes sparser, resulting in a simpler first-stage landscape. This trend should therefore not be assumed to persist for the considered classical solvers in general.

Across the tested values of $m$ and $|E|$, QASCO maintains its hit rate and a comparable ERT advantage over the classical methods, despite substantial variation in the calibrated SAA batch size and absolute query expenditure (see \cref{sec:appendix_full_solver_comparison_table}). The QASCO mixing mechanism coherently aggregates scenario and recourse information so that its first-stage mixer acts on one reference amplitude per first-stage decision. Increasing either $m$ or $|E|$ at fixed $n$ enlarges the scenario space while leaving the first-stage and recourse decision spaces unchanged for stochastic MaxCut. The observed stability therefore provides evidence that QASCO's aggregation-based strategy remains effective as the scenario space grows.

We further observe in \cref{fig:solver-recourse-conditioned} that the QASCO hit rate remains stable as expected minimum recourse cost at the stochastic optimum increases, while the hit rate for each classical solver decreases substantially over the same range. Higher expected minimum recourse cost can reflect both finite repair requirements and scenarios in which no repair reaches the threshold. The decline in classical hit rates may partly reflect the increased number of queries required to compute $Q^*(x;\omega)$ for each scenario in the SAA batch (\cref{sec:appendix_evaluation_sampling_costs}). The classical solvers rely on SAA evaluations of candidates, so this additional cost limits the search space within the query budget. For QASCO, the additional recourse evaluation cost is incurred when evaluating candidates returned by the circuit but, at fixed $(p_R,p)$, it does not change the number of queries per circuit preparation.

Our circuit analysis identifies implementation of the recourse oracle as the primary resource challenge. Under the assumptions in \cref{sec:maxcut_implementation}, the per-layer two-qubit gate complexity is $O((1+p_R)[|E|(b+\log|E|)+n])$, with repeated recourse-oracle applications dominating the cost for fixed $p_R$. Explicit recourse encoding requires an $n$-qubit register, giving $2n+n_\Omega$ non-ancilla qubits in total. Nevertheless, this polynomial per-layer resource scaling is sufficient for the ansatz to act on a configuration space whose size grows exponentially with the number of encoded variables.

The two-stage QAOA uses $O(n)$ single-qubit gates for each transverse-field mixing step. The QASCO controlled mixer requires $O(n+n_\Omega)$ two-qubit gates, while its associated basis transformations additionally require the scenario and recourse preparation circuits and their inverses. With at most a fixed number of perturbation values per edge and $|E|=\Theta(n)$, the controlled mixer has $O(n)$ gate complexity; repeated recourse-oracle applications add $O((1+p_R)n(b+\log n))$ per layer. The additional controlled operations and oracle calls may make QASCO harder to implement on noisy intermediate-scale quantum devices. Nevertheless, for the problems studied here, our results suggest that QASCO is likely to be preferable to the two-stage QAOA when adequate quantum hardware is available. The original two-stage QAOA proposal used a trained quantum generative adversarial network (qGAN) to prepare the scenario distribution~\cite{kuroiwa2025}. Such a method could also implement $S_\Omega$ in QASCO when efficient for the target distribution~\cite{Zoufal2019qGAN}.

While we optimise the schedule parameters by exact evaluation of $f_\mathrm{joint}$ here, the calibrated parameters show consistent trends with ansatz depth (see \cref{sec:appendix_linear_ramp_schedule_parameters}). Parameter transfer from smaller to larger problem instances~\cite{GaldaEtAl2021QAOATransferability} or from shallower to deeper circuits~\cite{zhou_quantum_2020} may therefore lower the cost of optimisation on hardware in practice. Fully shot-based parameter optimisation and optimisation in the presence of noise remain key areas for future investigation.

The successful use of the transverse-field mixer within QASCO suggests a route to stochastic optimisation problems whose first-stage search spaces admit existing QAOA or QWOA constructions. The first-stage mixing graph determines which first-stage decisions can couple, while the reference-state amplitudes determine the associated couplings over the scenario and recourse registers. Existing QAOA or QWOA state-preparation and mixing constructions that preserve feasibility~\cite{hadfield2019quantum,FuchsEtAl2022ConstraintMixers} could therefore support extensions to constrained first-stage search spaces.

\section{Conclusion}
\label{sec:conclusion}

We introduced Quantum Approximate Stochastic Combinatorial Optimisation (QASCO), a quantum approach to two-stage stochastic optimisation with recourse. QASCO coherently combines information about possible scenarios and recourse actions to increase the probability of measuring first-stage decisions with low cost under the stochastic objective, without explicitly evaluating their expected minimum recourse costs. We apply it to a MaxCut problem with stochastic edge weights, constructing a circuit whose gate count per layer grows polynomially with the number of problem variables.

In exact state-vector simulations, QASCO more consistently converges to first-stage decisions optimal under the stochastic objective than a previously proposed QAOA-based method under comparable parameter-optimisation effort. We apply a fixed-parameter-count schedule originally proposed for deterministic quantum approximate optimisation to QASCO and find that parameters calibrated on a set of representative instances can amplify low-cost states when applied to new problem instances without per-instance parameter optimisation. Compared to classical uniform enumeration, greedy search, and tabu search using sample average approximation of the stochastic objective, QASCO consistently reached a global optimum in the benchmarks, with a lower expected query cost under the stated model, and maintained its effectiveness across the tested increases in configuration-space size and expected minimum recourse cost at the stochastic optimum.

Studying QASCO under noise and developing end-to-end, shot-efficient parameter optimisation are natural next steps. An analytical investigation is also needed to identify the regimes in which QASCO is expected to converge effectively. Applying QASCO to additional stochastic combinatorial problems, particularly those with constrained first-stage and recourse decision spaces, will help establish its utility as a general framework for stochastic combinatorial optimisation.

\section*{Acknowledgements}
This project was funded through the Department of Defence Advanced Strategic Capabilities Accelerator (ASCA). Computational resources were provided by the \href{https://doi.org/10.48569/18sb-8s43}{Setonix supercomputer} system at the Pawsey Supercomputing Research Centre, which is funded by the Australian Government and the Government of Western Australia. Additional support came from Pawsey's Quantum Supercomputing Innovation Hub, made possible by a grant from the Australian Government through the National Collaborative Research Infrastructure Strategy (NCRIS). EM also thanks the Pawsey quantum team for their support and helpful discussions throughout this work.

\section*{Author contributions}
EM developed the proposed quantum walk-based optimisation method, devised and performed the numerical simulations, and prepared the initial manuscript draft. PS proposed the initial research direction, and provided expertise and guidance on problem formulation and solution approaches. JB supervised and guided the technical arguments, conclusions, and presentation of the manuscript. UV provided institutional support for the project and contributed to the manuscript's critical review. All authors contributed to writing, critically reviewing, and revising the manuscript. AI tools were used to assist with language editing, formatting and manuscript review.

\bibliographystyle{quantum}
\bibliography{bibliography}

\appendix

\FloatBarrier

\section{Independent uncertain parameters}
\label[appendix]{sec:appendix_independence_assumption}

Let $\omega_0, \ldots, \omega_{d_\omega-1}$ be mutually independent uncertain parameters. For each $i \in \{0, \ldots, d_\omega-1\}$, let $\Omega_i$ denote the support of $\omega_i$ and let $\mu_i(a) = \mathbb{P}\!\left[\omega_i = a\right]$, for $a\in\Omega_i$, be its marginal distribution. The scenario space factorises as $\Omega = \Omega_0 \times \cdots \times \Omega_{d_\omega-1}$,
and the joint distribution satisfies
\begin{equation*}
    \mu(\omega) = \prod_{i=0}^{d_\omega-1}\mu_i(\omega_i),
    \qquad \omega = (\omega_0, \ldots, \omega_{d_\omega-1}) \in \Omega.
\end{equation*}
In particular, a general distribution on $\Omega$ requires $|\Omega|-1$ free probability parameters, whereas the above product-form model requires only $\sum_{i=0}^{d_\omega-1}(|\Omega_i|-1)$. If each uncertain parameter has support satisfying $2\le |\Omega_i|\le\kappa$ for a constant $\kappa$ independent of $d_\omega$, then $\log|\Omega|=\Theta(d_\omega)$ and the product model requires $O(d_\omega)$ probability parameters.
 
\section{Contributions to the ansatz objective}
\label[appendix]{sec:appendix_objective_contributions}

For the $n=6$, $m=3$, $|E|=6$ instances considered in \cref{sec:comparison-with-two-stage_qaoa}, we write $\ket{\psi}=\ket{\psi_R}+\ket{\psi_\perp}$, with reference component $\ket{\psi_R}=\sum_x b_x\ket{x,r_x}$ and $\ket{\psi_\perp}$ orthogonal to $\mathcal S$. The squared projection norms of each final depth-$p$ state are $P_R=\sum_x|b_x|^2$ and $P_\perp=1-P_R$.

\begin{figure}[h]
    \centering
    \includegraphics[width=0.8\linewidth]{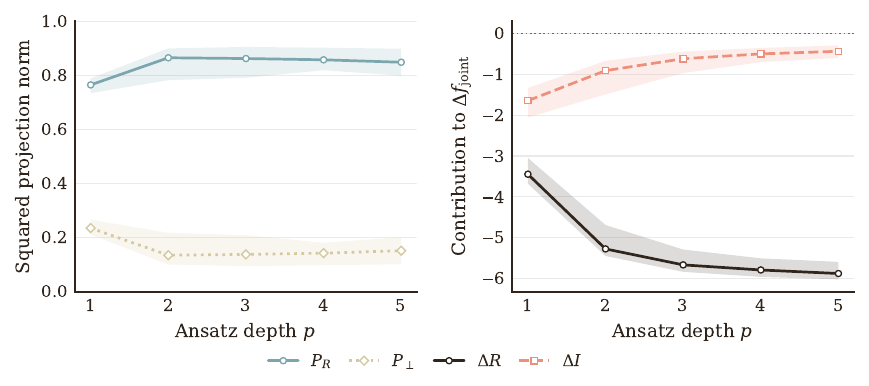}
    \caption{Reference-subspace occupation and objective contributions across the three rank-gap ensembles. \textbf{Left}: squared projection norms $P_R$ and $P_\perp$ of the final state. \textbf{Right}: cumulative changes $\Delta R_p$ and $\Delta I_p$ in the reference and interference contributions, where negative values reduce $f_{\mathrm{joint}}$. Lines show medians over the $150$ instances and bands show interquartile ranges.}
    \label{fig:appendix-objective-contributions}
\end{figure}

With respect to the cost operator $\hat H$ (see \cref{sec:variational_objective}), the reference and interference contributions are
\begin{equation*}
R:=\bra{\psi_R}\hat H\ket{\psi_R}
=\sum_x|b_x|^2F_s^\mu(x),\qquad
I:=2\operatorname{Re}\bra{\psi_R}\hat H\ket{\psi_\perp}.
\end{equation*}
We compute the cumulative changes $\Delta R_p$ and $\Delta I_p$ in these contributions over the $p$ mixing steps. \Cref{fig:appendix-objective-contributions} summarises these quantities across the three rank-gap ensembles. The median $P_R$ increases from $0.765$ at $p=1$ to $0.865$ at $p=2$ and is $0.849$ at $p=5$.

\section{Calibrated schedule parameters}
\label[appendix]{sec:appendix_linear_ramp_schedule_parameters}

\Cref{tab:linear-ramp-schedule-parameters} reports the selected QASCO depths and calibrated linear-ramp parameters for the eight problem settings in the candidate-generator benchmark of \cref{sec:solver_benchmark}. Both schedules use \cref{eq:linear_ramp}, except at $p_R=1$, where the recourse phase and walk times are $\gamma_{R,\max}$ and $t_{R,\max}$ and $\lambda_R$ is inactive. These parameters are fixed after calibration and used for all test instances in each setting.

\begin{figure}
    \centering
    \includegraphics[width=\linewidth]{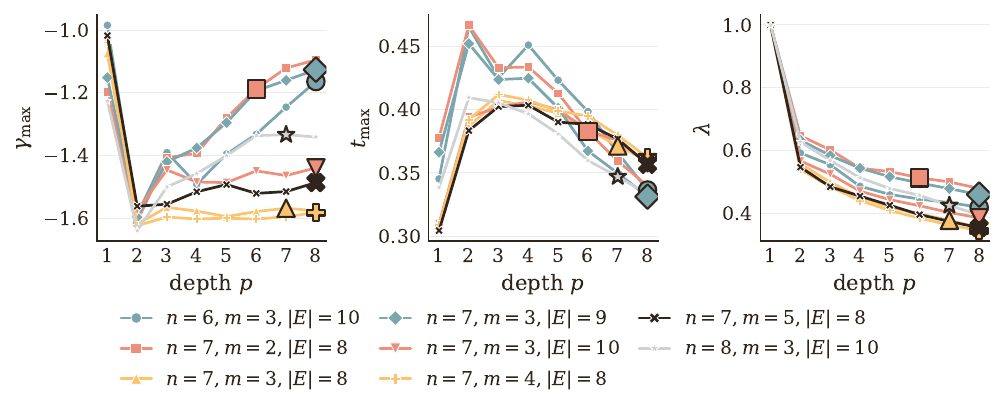}
    \caption{Calibrated ansatz linear-ramp parameters versus ansatz depth $p$ for the eight problem settings. \textbf{Left}: $\gamma_{\max}$. \textbf{Middle}: $t_{\max}$. \textbf{Right}: $\lambda$. For each curve, the recourse-preconditioner depth $p_R$ is fixed at the calibration-selected value (see \cref{sec:appendix_linear_ramp_schedule_parameters}). Larger outlined markers indicate the selected ansatz depth.}
    \label{fig:appendix-linear-ramp-schedule-parameters}
\end{figure}

\Cref{fig:appendix-linear-ramp-schedule-parameters} shows the calibrated ansatz parameters across $p=1,\ldots,8$, holding $p_R$ at its selected value for each setting. From $p=2$ to $p=8$, $\lambda$ decreases monotonically in every setting, from $0.539$--$0.647$ to $0.344$--$0.478$. The mixer parameter $t_{\max}$ reaches its maximum at $p=2$, $3$, or $4$ and decreases monotonically from $p=4$ onwards in every setting, reaching $0.331$--$0.362$ at $p=8$. The phase parameter $\gamma_{\max}$ is most negative at $p=2$ in every setting, but its subsequent variation differs between settings. At $p=8$ it ranges from $-1.582$ to $-1.095$. The selected ansatz depths lie between $6$ and $8$. 

\input{figures/linear_ramp_schedule_parameters_table.tex}

\section{Candidate-generator results}
\label[appendix]{sec:appendix_full_solver_comparison_table}

\Cref{tab:appendix-solver-comparison} reports hit rate and ERT with $95\%$ confidence intervals for the eight problem settings in the candidate-generator benchmark of \cref{sec:solver_benchmark}, together with the query budget $B_{\mathrm q}$, SAA batch size $M_0$ and selected tabu tenures. QASCO depths and parameter schedules are reported in \cref{sec:appendix_linear_ramp_schedule_parameters}.

\input{figures/solver_comparison/solver_comparison_appendix_table.tex}

\section{Recourse evaluation cost and SAA batch sizes}
\label[appendix]{sec:appendix_evaluation_sampling_costs}

\begin{figure}
    \centering
    \includegraphics[width=\linewidth]{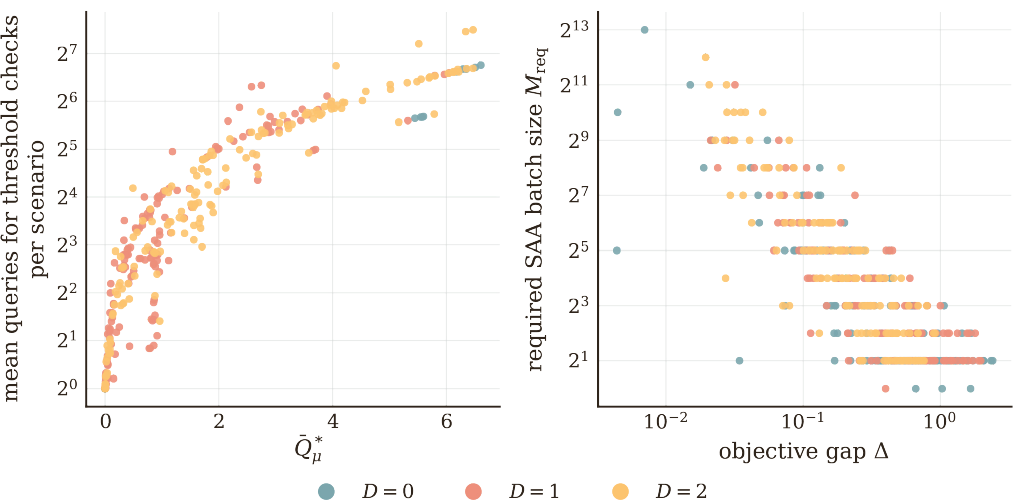}
    \caption{Recourse evaluation cost and required SAA batch sizes across the eight benchmark settings. \textbf{Left}: mean queries per sampled scenario for threshold checks at the stochastic optimum versus expected minimum recourse cost $\bar Q_\mu^*$, for the $480$ test instances. \textbf{Right}: required SAA batch size $M_{\mathrm{req}}$ versus objective gap $\Delta$, for the $480$ calibration instances. Colours indicate the repair-depth bins $D=0,1,2$ defined in \cref{sec:benchmark_ensembles}.}
    \label{fig:sop-difficulty-relationships}
\end{figure}

\Cref{fig:sop-difficulty-relationships} examines recourse evaluation cost and required SAA batch sizes for the calibration and test ensembles used in the candidate-generator benchmark of \cref{sec:solver_benchmark}.

For each test instance, we use the benchmark evaluator of \cref{sec:solver_comparison_protocol} to count the queries for threshold checks at the stochastic optimum, averaged over the scenarios in its five SAA batches. Each query evaluates a cut and tests whether its realised value meets or exceeds the threshold $T$. This measures per-scenario recourse evaluation cost before reuse of duplicate scenarios, excluding first-stage evaluation and candidate generation. Higher expected minimum recourse cost is associated with threshold-check queries, with mean counts ranging from $1$ to $180.6$. This relationship is consistent with the lower classical hit rates in the highest expected-cost group of \cref{fig:solver-recourse-conditioned}.

For each calibration instance, the objective gap $\Delta$ is the amount by which the stochastic objective $J_\mu$ of the best nonoptimal canonical solution exceeds the optimal value. We obtain the required SAA batch size $M_{\mathrm{req}}$ for each instance using the SAA calibration criterion in \cref{sec:solver_comparison_protocol}, then aggregate these into the calibrated batch size $M_0$ for each setting. Larger objective gaps require smaller SAA batch sizes. 

\end{document}

%% file: figures/linear_ramp_schedule_parameters_table.tex
\begin{table}
    \centering
    \caption{Calibration-selected QASCO configuration for each problem setting: ansatz depth $p$ and schedule $(\gamma_{\max}, t_{\max}, \lambda)$, and recourse-preconditioner depth $p_R$ and schedule $(\gamma_{R,\max}, t_{R,\max}, \lambda_R)$. Dashes indicate that no recourse preconditioner was selected ($p_R=0$).}
    \label{tab:linear-ramp-schedule-parameters}
    \small
    \begin{tabular}{lrrrrrrrr}
        \toprule
        Problem setting & $p$ & $\gamma_{\max}$ & $t_{\max}$ & $\lambda$ & $p_R$ & $\gamma_{R,\max}$ & $t_{R,\max}$ & $\lambda_R$ \\
        \midrule
        $n=6, m=3, |E|=10$ & 8 & -1.164 & 0.337 & 0.418 & 0 & -- & -- & -- \\
        $n=7, m=2, |E|=8$ & 6 & -1.188 & 0.383 & 0.512 & 0 & -- & -- & -- \\
        $n=7, m=3, |E|=8$ & 7 & -1.568 & 0.371 & 0.377 & 8 & -1.376 & 0.305 & 0.658 \\
        $n=7, m=3, |E|=9$ & 8 & -1.126 & 0.331 & 0.459 & 0 & -- & -- & -- \\
        $n=7, m=3, |E|=10$ & 8 & -1.440 & 0.358 & 0.385 & 4 & -1.526 & 0.339 & 0.799 \\
        $n=7, m=4, |E|=8$ & 8 & -1.582 & 0.362 & 0.344 & 8 & -1.330 & 0.305 & 0.641 \\
        $n=7, m=5, |E|=8$ & 8 & -1.486 & 0.357 & 0.353 & 8 & -1.248 & 0.305 & 0.649 \\
        $n=8, m=3, |E|=10$ & 7 & -1.333 & 0.347 & 0.424 & 1 & -1.456 & 0.298 & 0.999 \\
        \bottomrule
    \end{tabular}
\end{table}

%% file: figures/solver_comparison/solver_comparison_appendix_table.tex
\begin{center}
\small
\setlength{\tabcolsep}{2.5pt}
\begin{longtable}{@{}llrrrr@{}}
\caption{Candidate-generator benchmark results. QASCO uses the calibration-selected depths and schedules in \cref{sec:appendix_linear_ramp_schedule_parameters}; tabu search uses the calibration-selected tenure $\tau$. Each method has $300$ runs per setting ($60$ instances $\times$ five runs), or $2400$ overall; hit rate and ERT include unsuccessful runs. The $95\%$ percentile bootstrap confidence intervals are reported as relative percentage offsets. Hit-rate intervals are omitted when every run succeeded.}\label{tab:appendix-solver-comparison}\\
\toprule
Setting $(n,m,|E|)$ & Method & Hit rate & ERT (queries) & $B_{\mathrm q}$ & $M_0$ \\
\midrule
\endfirsthead
\toprule
Setting $(n,m,|E|)$ & Method & Hit rate & ERT (queries) & $B_{\mathrm q}$ & $M_0$ \\
\midrule
\endhead
\midrule
\multicolumn{6}{r}{\emph{continued on next page}}\\
\endfoot
\bottomrule
\endlastfoot
$(6,3,10)$ & QASCO & $1.000$ & $127710\,[-29.2\%,+30.4\%]$ & 1416256 & 4096 \\
 & Tabu ($\tau=3$) & $0.860\,[-7.8\%,+7.3\%]$ & $763187\,[-17.2\%,+21.4\%]$ &  &  \\
 & Greedy & $0.853\,[-7.7\%,+7.0\%]$ & $809119\,[-17.0\%,+20.1\%]$ &  &  \\
 & Uniform & $0.780\,[-9.9\%,+9.9\%]$ & $997056\,[-17.5\%,+21.1\%]$ &  &  \\
\addlinespace[0.25em]
$(7,2,8)$ & QASCO & $1.000$ & $3498\,[-22.5\%,+25.9\%]$ & 58758 & 64 \\
 & Tabu ($\tau=4$) & $0.903\,[-5.9\%,+5.2\%]$ & $24362\,[-16.6\%,+19.1\%]$ &  &  \\
 & Greedy & $0.910\,[-5.8\%,+4.7\%]$ & $23390\,[-16.1\%,+21.1\%]$ &  &  \\
 & Uniform & $0.813\,[-8.6\%,+8.2\%]$ & $38049\,[-16.2\%,+19.9\%]$ &  &  \\
\addlinespace[0.25em]
$(7,3,8)$ & QASCO & $1.000$ & $53669\,[-33.9\%,+38.8\%]$ & 1043547 & 1024 \\
 & Tabu ($\tau=3$) & $0.953\,[-3.8\%,+3.6\%]$ & $366599\,[-15.0\%,+16.8\%]$ &  &  \\
 & Greedy & $0.930\,[-5.7\%,+5.1\%]$ & $408522\,[-17.1\%,+19.7\%]$ &  &  \\
 & Uniform & $0.833\,[-7.6\%,+7.2\%]$ & $629430\,[-14.6\%,+18.5\%]$ &  &  \\
\addlinespace[0.25em]
$(7,3,9)$ & QASCO & $1.000$ & $35569\,[-28.9\%,+34.1\%]$ & 777398 & 512 \\
 & Tabu ($\tau=1$) & $0.923\,[-5.0\%,+4.3\%]$ & $289528\,[-17.4\%,+21.0\%]$ &  &  \\
 & Greedy & $0.890\,[-6.4\%,+6.0\%]$ & $334326\,[-18.3\%,+23.0\%]$ &  &  \\
 & Uniform & $0.833\,[-7.6\%,+6.8\%]$ & $469639\,[-15.2\%,+19.5\%]$ &  &  \\
\addlinespace[0.25em]
$(7,3,10)$ & QASCO & $1.000$ & $61476\,[-30.1\%,+34.5\%]$ & 1467495 & 1024 \\
 & Tabu ($\tau=1$) & $0.937\,[-4.7\%,+3.5\%]$ & $506957\,[-16.2\%,+19.0\%]$ &  &  \\
 & Greedy & $0.937\,[-4.3\%,+3.8\%]$ & $544783\,[-16.0\%,+18.2\%]$ &  &  \\
 & Uniform & $0.850\,[-6.7\%,+6.7\%]$ & $842200\,[-14.9\%,+17.5\%]$ &  &  \\
\addlinespace[0.25em]
$(7,4,8)$ & QASCO & $1.000$ & $415740\,[-33.3\%,+38.0\%]$ & 9527076 & 8192 \\
 & Tabu ($\tau=4$) & $0.970\,[-2.8\%,+2.1\%]$ & $3115811\,[-13.4\%,+14.0\%]$ &  &  \\
 & Greedy & $0.950\,[-3.9\%,+3.5\%]$ & $3232464\,[-14.0\%,+15.3\%]$ &  &  \\
 & Uniform & $0.873\,[-5.7\%,+5.0\%]$ & $5588382\,[-11.5\%,+13.3\%]$ &  &  \\
\addlinespace[0.25em]
$(7,5,8)$ & QASCO & $1.000$ & $91350\,[-26.2\%,+31.4\%]$ & 1699004 & 1024 \\
 & Tabu ($\tau=3$) & $0.933\,[-4.6\%,+4.0\%]$ & $661872\,[-17.1\%,+20.2\%]$ &  &  \\
 & Greedy & $0.923\,[-4.3\%,+4.0\%]$ & $662992\,[-16.1\%,+19.2\%]$ &  &  \\
 & Uniform & $0.797\,[-8.4\%,+7.9\%]$ & $1137803\,[-15.1\%,+18.2\%]$ &  &  \\
\addlinespace[0.25em]
$(8,3,10)$ & QASCO & $1.000$ & $33762\,[-36.4\%,+46.3\%]$ & 1238403 & 256 \\
 & Tabu ($\tau=4$) & $0.967\,[-3.8\%,+2.7\%]$ & $300002\,[-18.3\%,+22.7\%]$ &  &  \\
 & Greedy & $0.940\,[-4.6\%,+3.9\%]$ & $347394\,[-19.4\%,+24.1\%]$ &  &  \\
 & Uniform & $0.840\,[-7.5\%,+7.5\%]$ & $772833\,[-15.0\%,+16.8\%]$ &  &  \\
\end{longtable}
\end{center}